\documentclass[12pt,preprint,numberedappendix]{emulateapj}
\usepackage{lscape}

\begin{document}

\title{The Coevality of Young Binary Systems}

\author{Adam L. Kraus\altaffilmark{1} (alk@astro.caltech.edu), Lynne A. Hillenbrand\altaffilmark{1}}
\altaffiltext{1}{California Institute of Technology, Department of 
Astrophysics, MC 105-24, Pasadena, CA 91125}

\begin{abstract}

Multiple star systems are commonly assumed to form coevally; they thus 
provide the anchor for most calibrations of stellar evolutionary models. In 
this paper we study the binary population of the Taurus-Auriga association, 
using the component positions in an HR diagram in order to quantify the 
frequency and degree of coevality in young binary systems. After identifying 
and rejecting the systems that are known to be affected by systematic errors 
(due to further multiplicity or obscuration by circumstellar material), we 
find that the relative binary ages, $|\Delta$$\log\tau|$, have an overall 
dispersion $\sigma$$_{|\Delta\log\tau|}$$\sim$0.40 dex. Random pairs of 
Taurus members are coeval only to within 
$\sigma$$_{|\Delta\log\tau|}$$\sim$0.58 dex, indicating that Taurus binaries 
are indeed more coeval than the association as a whole. However, the 
distribution of $|\Delta$$\log\tau|$ suggests two populations, with 
$\sim$2/3 of the sample appearing coeval to within the errors 
($\sigma$$_{|\Delta\log\tau|}$$\sim$0.16 dex) and the other $\sim$1/3 
distributed in an extended tail reaching $|\Delta$$\log\tau|$$\sim$0.4-0.9 
dex. To explain the finding of a multi-peaked distribution, we suggest that 
the tail of the differential age distribution includes unrecognized 
hierarchical multiples, stars seen in scattered light, or stars with disk 
contamination; additional followup is required to rule out or correct for 
these explanations. The relative coevality of binary systems does not depend 
significantly on the system mass, mass ratio, or separation. Indeed, any 
pair of Taurus members wider than $\sim$10\arcmin\, ($\sim$0.7 pc) shows the 
full age spread of the association.

\end{abstract}

\keywords{stars: binaries: general, stars: evolution, 
stars: fundamental parameters, stars: Hertzsprung-Russell diagram, 
stars: pre-main sequence, stars: open clusters and 
associations: individual(Taurus-Auriga)}

\section{Introduction}

Stellar evolutionary models are critical for interpreting astronomical 
observations, but they are not well-calibrated for pre-main sequence (PMS) 
stars. Such calibration requires the measurement of some or all of the 
fundamental stellar properties: age, mass, radius, luminosity, and 
effective temperature. Ages are notoriously difficult to estimate (Mamajek 
\& Hillenbrand 2008; Hillenbrand 2009), though they can be inferred 
indirectly from a membership in a stellar population for which the mean 
age can be determined. Stellar masses and/or radii require orbital 
monitoring of eclipsing or visual binary systems. The known pre-main 
sequence eclipsing binary systems sparsely sample parameter space due to 
their extreme rarity (e.g. Irwin et al. 2007; Stassun et al. 2007; 
Stempels et al. 2008). Most PMS visual binaries have only partial 
orbits because young stars are distant and any systems which can be 
spatially resolved necessarily have wide separations and corresponding 
long periods (Steffen et al. 2001; Duch\^ene et al. 2006), though a 
handful of short-period systems are bright enough for interferometric 
techniques to be feasible (Boden et al. 2005; Schaefer et al. 2008). In 
contrast to ages, masses, and radii, the luminosities and temperatures of 
stars are straightforward to infer from single-epoch observations, so they 
offer the best near-term prospects for systematic calibration of stellar 
models. The procedure can also be inverted: given a star's luminosity and 
temperature, a theoretical model can be used to estimate its age and mass, 
plus its radius can be estimated directly from the Stefan-Boltzmann law.

The standard procedure for calibrating models with luminosities and 
temperatures is to place two or more nominally coeval stars on an HR 
diagram. These stars should trace an empirical isochrone sequence, and 
this sequence can be compared to theoretical isochrones in order to test 
their consistency with observations. HR diagram analysis has provided 
many crucial insights into models of stellar interiors, atmospheres, and 
evolution (e.g. White et al. 1999; Luhman et al. 2003). There are many 
systematic astrophysical effects that can complicate this analysis, 
including unresolved multiplicity, obscuration from circumstellar 
material (i.e. an envelope or edge-on disk), and veiling from accretion 
(at blue optical wavelengths) or circumstellar disk emission (at 
near-infrared wavelengths). These effects can yield ages with errors of 
an order of magnitude or more, so the samples used in this analysis must 
be inspected closely to identify stars that might be affected. 
Additional physics, such as stellar activity or convection efficiency, 
could also play a role (Chabrier et al. 2007; Stassun et al. 2008).

The inverse procedure plays an important role in the study of binary 
systems (e.g. Hartigan et al. 1994; White \& Ghez 2001; Hartigan \& Kenyon 
2003). It is commonly assumed that multiple star formation proceeds almost 
simultaneously, such that all stars in a bound multiple system are coeval. 
However, this assumption can be tested only by using the evolutionary 
models that require calibration, so any apparent disagreement between 
binary component ages could be due to non-coevality or errors in the 
models. Emerging evidence from several young eclipsing binary systems 
shows that their component properties are inconsistent with any single age 
predicted by pre-main sequence evolutionary models (e.g. Stassun et al. 
2007, 2008), but the frequency and degree of implied noncoevality is still 
unclear.

In this paper, we estimate the ages of a large sample of 
stringently vetted young binary systems in the Taurus association 
($\tau\sim$1-2 Myr; $d\sim$145 pc) in order to test the system 
components' relative coevality and the validity of theoretical 
isochrones in matching empirical HR diagram sequences. In Section 
2, we describe our sample of binary systems, and in Section 3, we 
describe the stellar models and analysis techniques used to infer 
stellar ages. In Section 4, we show an HR diagram with all our 
sample members and identify likely contaminants. In Section 5, we 
test the coevality of young binary systems by adopting 
model-predicted ages. Finally, in two appendices, we test the 
evolutionary models using a large sample of likely single stars and 
using a subset of high-order multiple systems.

\section{The Sample}

The accurate determination of stellar parameters requires spatially- and/or 
spectrally-resolved observations that are not polluted by light from 
companions; our sample is comprised of all known Taurus binary systems that 
have spectral types and flux ratios for at least two components. At the 
distance of Taurus ($\sim$145 pc), most binary systems with separations of 
$\la$100--200 AU can not be resolved in seeing-limited observations with 
typical resolutions of $\sim$1\arcsec. Given this limit, most sample members 
have been observed with HST, ground-based adaptive optics, or echelle 
spectrographs (distinguishing the components' spectra and relative fluxes). 
High-order multiplicity can give the appearance of non-coevality, so we 
immediately omitted any binary pairs which contained an additional component 
without its own measured spectral type and flux. As we describe in Section 
4, we also cut additional targets from this sample if they appeared to be 
affected by other systematic uncertainties. We list the observed properties 
of our sample members in Table 1, along with the references used to infer 
spectral types, extinctions, and fluxes. Given the ad hoc nature of our 
sample, it is not complete and may be subject to biases, but the goal of our 
initial sample selection was to be as inclusive as possible.
 
The wide components that could be resolved in seeing-limited observations 
were drawn from the sample studied in our previous wide multiplicity 
survey (Kraus \& Hillenbrand 2007a, 2009). Most of these stars had already 
been identified as Taurus members, so we drew their observed properties 
from previous work by Kenyon \& Hartmann (1995), Duch\^ene et al. (1999), 
White \& Basri (2003), White \& Hillenbrand (2004), and Luhman (2004, 
2006). All of these authors reported spectral types and extinctions, and 
we adopted NIR magnitudes from the 2MASS Point Source Catalog (Skrutskie 
et al. 2006) or from our own PSF fitting photometry of 2MASS atlas images 
(Kraus \& Hillenbrand 2007a). The 2MASS photometry for RW Aur AB and most 
of the Duch\^ene et al. sample was unreliable since the system separations 
fell near the 2MASS resolution limits, so we adopted total $K$ fluxes from 
2MASS and the $K$ band flux ratios reported by White \& Ghez (2001) or 
Correia et al. (2006). We adopted an outer separation limit of 30\arcsec\, 
for identifying binary pairs. The binary population probably dominates 
among all pairs extending out to $\sim$2\arcmin\, (Kraus \& Hillenbrand 
2008), but the frequency of chance alignments between unbound stars 
becomes significant at $>$30\arcsec.

The closer systems that could be resolved with high-resolution imaging 
were drawn from several recent spectroscopic surveys. Most of the 
spectroscopic observations were obtained with HST/STIS by Hartigan \& 
Kenyon (2003), but several systems were observed under good seeing by 
Duch\^ene et al. (1999). Individual systems were also studied by White et 
al. (1999) with HST/FOS, or in our own survey of low-mass multiplicity 
with Keck laser guide star AO (LGSAO; Kraus et al., in prep). As before, 
we inferred NIR magnitudes from the total system fluxes reported in 2MASS 
and spatially resolved flux ratios reported by Leinert et al. (1993), 
White \& Ghez (2001), Correia et al. (2006), and our LGSAO survey.

There are a small number of double-lined spectroscopic binaries that have 
been identified and studied in some detail in Taurus. We have added the 
well-known system UZ Tau Aab to our sample, adopting the spectral types 
and inferred H band flux ratio found by Prato et al. (2002). We also added 
the short-period spectroscopic binary V773 Tau Aab, which was studied with 
RV and interferometric monitoring by Boden et al. (2007) as part of an 
orbit monitoring program. The V773 Tau system also includes two faint 
companions at small separations and a wide brown dwarf companion; we do 
not include the close companions because they do not have spectral type 
determinations and because they are too faint to influence the observed 
properties of V773 Tau Aab (Duch\^ene et al. 2003), but we include the 
wide substellar companion (Luhman et al. 2004). We also considered whether 
to include V826 Tau (Massarotti et al. 2005) and DQ Tau (Mathieu et al. 
1997), but the only known flux ratio for V826 Tau is in the optical and 
there are no flux ratios reported for DQ Tau, so these systems could not 
be integrated with the rest of our sample.

We also note that one binary system in our sample, XZ Tau, was recently 
suggested to be a possible hierarchical triple based on resolved radio 
observations of XZ Tau B at 1.3 cm and 7 mm. We have retained this 
system in our sample because the companion was not identified with HST 
(Krist et al. 2008) or in other surveys, but the suggested separation 
($\sim$90 mas or 13 AU) was near the detection limits of the earlier 
observations, so we regard it as a credible possibility. As we will show 
later, XZ Tau B is overluminous compared to XZ Tau A or HL Tau, which 
supports its probable multiplicity.

Finally, most of the binary components in our sample have spatially 
resolved photometry in the $K$ filter only. Inferring the component 
luminosities from such a red bandpass might introduce systematic 
errors in our estimated luminosities due to near-infrared excesses 
from circumstellar disks. In order to address this prospect, we have 
searched the literature to determine which stars are likely to host a 
warm disk; we summarize our assessments and the corresponding 
references in Table 1. We based these assessments, in order of 
priority, on 3-10 $\mu$m photometry (from Spitzer/IRAC or ground-based 
AO imaging), optical spectroscopic accretion signatures, 10-30 $\mu$m 
spectroscopy (from Spitzer/IRS), and finally on submm/mm photometry. 
In each case where sufficient data is available, we have concluded 
that the star either has a disk (``Y''), does not have a disk (``N''), 
or might have a disk (but the observations aren't spatially resolved, 
so we can't determine which binary component(s) have one; ``Y?''). We 
will address the significance of flux excesses from warm dust in 
Section 5.

\section{Analysis}

\subsection{Inferred and Calculated Stellar Properties}

Any comparison of observations to stellar evolutionary models requires the 
conversion of observed properties (spectral types, filtered magnitudes, and 
extinctions) into fundamental physical parameters (effective temperatures 
and bolometric luminosities). This process is accomplished by invoking a 
temperature scale, a set of bolometric corrections, a reddening law, and an 
estimated distance. Temperature scales directly relate spectral types to 
temperatures, allowing the estimation of temperatures based on observed 
spectral features. Temperatures are also used to define intrinsic colors, 
from which an observed color can be used to infer the reddening and 
extinction. Bolometric corrections are temperature-dependent ratios of the 
flux in a filtered band to the full bolometric flux and are calibrated for 
nearby field stars that have been studied across the full range of 
wavelengths with significant contribution to the luminosity. Once the 
bolometric flux is known, the distance for a star then directly yields the 
bolometric luminosity.

Observed spectral types can be converted into effective temperatures using 
an adopted temperature scale, but this process is somewhat uncertain for 
young stars since the relation between temperature and spectral type is 
sensitive to surface gravity. For M-type stars, a giant with a given 
spectral type can be $\sim$300 K warmer than a dwarf of the same type 
(e.g. Leggett et al. 1996 versus Perrin et al. 1998; Richichi et al. 1998; 
van Belle et al. 1999). Young stars have intermediate 
surface gravity, so it has been suggested (e.g. Luhman 1997; Luhman et al. 
2003) that an intermediate temperature scale might be appropriate. For 
example, Luhman et al. (2003) found that M2-M5 stars in IC 348 have bluer 
$R-I$ colors than their field counterparts, matching the trend seen for 
giants. Similarly, we have shown in a previous publication (Kraus et al. 
2006) that the $V-I$ colors of M5-M9 stars in Taurus and Upper Sco are 
significantly bluer than those of field dwarfs, though not as blue as for 
giants. Conversely, more recent surveys of moderately older stars in 
$\epsilon$ Cha and $\eta$ Cha by Lyo et al. (2004, 2008) show that the 
broadband colors and temperature-sensitive narrowband spectral indices are 
generally similar to field dwarfs to spectral types as late as M5.5, with 
only a mild discrepancy in the $B-V$ color that could be attributed to 
chromospheric activity.

We have adopted the temperature scale suggested by Luhman et al. (2003) 
for use in low-gravity young stars. For spectral types $\la$M0, the 
temperature-SpT relation does not appear to be gravity-sensitive, so 
Luhman et al. use the temperature scale of Schmidt-Kaler (1982). For 
late-type ($\ga$M0) young stars, Luhman et al. define an intermediate 
temperature scale that makes the average cluster sequences of Taurus and 
IC348 internally coeval with respect to the NextGen models (Baraffe et 
al. 1998). Unfortunately, this calibration introduces an element of 
circularity into our analysis; the temperature scale is chosen by 
definition to make our chosen models reproduce observations. It is 
unclear whether the temperature scale of young stars is truly different 
or the adopted temperature scale is correcting for a discrepancy in the 
model, so our results should be weighed accordingly. The difference in 
inferred temperatures with respect to field M dwarf values (Leggett et 
al. 1996) is $\sim$50 K at M3, $\sim$100 K at M5, and $\sim$200 K at M8, 
equivalent to a systematic uncertainty of $\pm$0.25, $\pm$0.5, and 
$\pm$0.75 subclass, respectively.

We have adopted the bolometric corrections that we previously described 
for use with field stars (Kraus \& Hillenbrand 2007b). For spectral types 
$\la$K7, we used the corrections suggested by Masana et al. (2006), while 
for M dwarfs, we used the corrections of Leggett et al. (1992) and Leggett 
et al. (1996). Values calibrated for field dwarfs might be systematically 
inaccurate if the bolometric correction is also gravity-sensitive, so it 
would be helpful to verify whether these bolometric corrections are valid 
for young stars. However, we are not aware of any such tests having been 
attempted. In all cases, we have adopted the observed NIR flux (and its 
bolometric correction) which is closest to the $J$ filter. As we describe 
below, luminosity excesses are typically least significant in the $I$ and 
$J$ filters, but most of the binary systems in our sample have 
resolved flux measurements only in the $H$ or $K$ filters.

We have implemented our extinction corrections using the interstellar 
reddening law of Schlegel et al. (1998), which stipulates that one 
magnitude of visual extinction corresponds to $A_J=0.28$, $A_H=0.18$, 
and $A_K=0.11$; these values are consistent with the interstellar 
reddening law in the 2MASS filters suggested by Indebetouw et al. 
(2005). Reddening laws might vary in regions with extremely high density 
(Weingartner \& Draine 2001; Rom\'an-Z\'u\~niga et al. 2007), but most 
of our sample members are only moderately reddened. Analysis of 2MASS 
source counts and colors toward the Ophiuchus, Lupus, and Pipe Nebulae 
suggest that the standard interstellar reddening law is appropriate to 
extinctions of $A_V$$\sim$20 or more (Lombardi et al. 2008).

We adopted a characteristic distance for all Taurus members of 
145$\pm$15 pc. Recent high-precision parallax measurements with 
the VLBA (Lestrade et al. 1999; Loinard et al. 2007; Torres et al. 
2007; Loinard et al. 2008) suggest that there might be a distance 
gradient of 165-125 pc in the east-west direction, though the 
discrepant distances of neighboring V773 Tau and Hubble 4 
(148$\pm$5 pc versus 132.5$\pm$0.6 pc) suggest an overall scatter 
at any location of $\sim$10-15 pc (10\%). The luminosity 
uncertainty if we adopt the characteristic distance is only 
$\sim$0.1-0.2 mag, which is similar to the uncertainty from 
dereddening and intrinsic variability, so attempting to 
extrapolating more precise distances from this suggestion of 3D 
structure is not likely to improve our results.

We should note that there is room for significant uncertainty in our 
inferred luminosities due to the intrinsic variability of young stars. 
Variability in Class III stars should be caused by cool spots, so its 
characteristic amplitude at near-infrared wavelengths should be no more 
than $\sim$0.1 mag (e.g. Carpenter et al. 2001, 2002). However, the same 
survey showed that Class I-II stars occasionally vary by as much as 1-2 
mag in the near infrared, and extreme classes of stars (such as FUor and 
EXor stars) can vary by even more. This suggests that an unusually young 
apparent age for one binary component could be the result of variability 
in that component.

A star that is surrounded by circumstellar material could also appear 
systematically underluminous by several magnitudes if it possesses an 
edge-on disk or a massive circumstellar envelope that renders it visisble 
only in scattered light. It would then appear much older than its 
unobscured companion. One such case is HK Tau B; both components of this 
system have a spectral type of M1 (White \& Hillenbrand 2004), but the B 
component is 3.3 magnitudes fainter in $K$. 

Finally, the presence of veiling due to accretion (at blue wavelengths) or 
circumstellar dust emission (at red wavelengths) could bias the inferred 
luminosity of a binary component. Observations in the $I$ or $J$ filter 
are least sensitive to this veiling, though even those measurements can be 
significantly impacted (Folha \& Emerson 1999; White \& Hillenbrand 2004). 
In our study, we have used photometry from $J$ or from the nearest redder 
filter. We would prefer to use $J$ for all systems, but in most cases, the 
absence of suitable data forces us to use relative photometry in $H$ or 
$K$. Data from bluer filters is also available for some sample members, 
but these data are not as homogeneous and we prefer to minimize the total 
wavelength range over which we make our luminosity estimates.

We list the stellar properties that we inferred with these methods in 
Table 1. Our temperature uncertainties have been determined from the 
uncertainty in each star's spectral type. Each of the bolometric magnitude 
uncertainties listed above should contribute $\sim$0.1 mag, so we have 
adopted a total statistical uncertainty of 0.3 mag. However, many of the 
uncertainties (such as for distance and extinction) should be correlated 
between binary components, so the uncertainty in their relative ages 
should be lower than our formal estimates. As we discuss further in 
Section 5, there is compelling evidence that the scatter in relative ages 
of binary components is indeed significantly smaller, suggesting that 
these error estimates are conservative.

\subsection{Inferred Physical Stellar Parameters}

Several sets of pre-main sequence evolutionary models have been 
developed in recent decades, but all of these models still face 
significant challenges in confronting observational constraints. 
Hillenbrand \& White (2004) found that all models have difficulty 
matching the dynamical masses of young stars with $M\la$1.2 
$M_{\sun}$, a range which encompasses almost all of our sample. 
However, the Lyon models (Baraffe et al. 1998; Chabrier et al. 
2000) seem to work best for low-mass stars, especially when using 
a mixing length of $\alpha$$=1.0$ and the revised temperature 
scale of Luhman et al. (2003). All models were found to reproduce 
the observed masses for stars with $M\ga$1.2 $M_{\sun}$, though 
the Lyon models only extend to 1.4 $M_{\sun}$.

In light of these results, we have adopted a hybrid combination of the 
Lyon models for low-mass stars and the models of D'Antona and Mazzitelli 
(1997; DM97) for higher-mass stars. For masses $\le$0.5 $M_{\sun}$, we 
use the mass-luminosity-temperature relations of the Lyon models with a 
mixing length of $\alpha$$=1.0$. For masses $\ge$1.0 $M_{\sun}$, we use 
the corresponding relations of DM97. Finally, in the intermediate regime 
of 0.6-0.9 $M_{\sun}$, we adopt a weighted average of the luminosity and 
temperature predicted by each model in order to produce a smooth 
transition between the two sets of models. This choice does not provide 
any insight into the missing physics that are still required to bring 
the models into agreement, but it represents an acceptable compromise 
for estimating relative ages of a sample of young stars. The two sets of 
models converge at older ages, so our solution is only important for 
very young stars ($\la$10 Myr).

These models report stellar luminosities and temperatures at 
quantized values of age and mass, so for each of our sample 
members, we have linearly interpolated between the four values of 
$T_{eff}$ and $M_{bol}$ around it in the HR diagram. The Lyon 
models also face a significant challenge with respect to very 
young stars since they are not defined for ages of $<$1 Myr, so 
for each star that falls above this isochrone and has a mass 
within the affected range ($<$1 $M_{\sun}$), we have linearly 
extrapolated its age from the four points below it in the HR 
diagram. These extrapolated ages should be regarded as much more 
uncertain than older ages, but the degree of error should be 
similar for stars with similar HR diagram positions, so only 
systems with disparate masses will be subject to the full 
systematic uncertainty.

In Table 2, we list the inferred mass and age for each star from our 
hybrid system and from the default Lyon and DM97 models. Several sample 
members illustrate the extreme difference in mass and age estimates for 
the two sets of models. For example, the inferred parameters of HN Tau A 
are $M=$1.35 $M_{\sun}$ and $\log(\tau)=$6.85 according to the Lyon models 
and $M=$0.65 $M_{\sun}$ and $\log(\tau)=$6.05 according to the DM97 
models; our hybrid isochrones yield $M=$0.85 $M_{\sun}$ and 
$\log(\tau)=$6.27. The uncertainty in $\log(\tau)$ due to observational 
errors is $\pm$0.25 dex for an average star in our sample, but as we 
describe below, this uncertainty could be an overestimate since binary 
systems appear to be more coeval.

\section{The HR Diagram}

The overall population sequence in an HR diagram provides a 
valuable test of our choice of evolutionary models. If the 
association is nominally coeval, then it should trace a single 
recognizable sequence that is parallel to theoretical isochrones. 
Individual sample members that strongly deviate from the 
association sequence should also be examined for a systematic 
source of error such as erroneous membership, circumstellar 
material that blocks and scatters the stellar flux, or 
misclassification of spectral types.

In Figure 1, we show the HR diagram for all of our binary sample 
members. The cluster sequence seems to trace the 1-2 Myr isochrone, 
albeit with significant scatter due to the many sources of uncertainty 
and the unknown spread of stellar ages. We adopted the two sets of 
theoretical isochrones (Lyon for low-mass stars and DM97 for higher-mass 
stars) specifically because they fit the Taurus single-star sequence 
across the full mass range of our sample (Appendix A), so this agreement 
with the canonical age of Taurus is not surprising. However, many 
individual members fall unusually low; as we show in Figure A2 for 
single stars, the 5 Myr isochrone represents a +2$\sigma$ deviation from 
the median age of Taurus, so we adopt this approximate lower limit for 
the observed scatter of the main body of members. Most of the binary 
components below this isochrone are known to be anomalous.

The warmest anomalous members (HBC 352-357) have been classified as Taurus 
members for several decades (e.g. Walter et al. 1988; Herbig \& Bell 1988; 
Kenyon \& Hartmann 1995) based on their X-ray emission and (for HBC 
352-355) their radial velocities. However, these stars are located at the 
far western edge of Taurus, well away from the central cloud cores, and 
their underluminosity has been recognized since their discovery. Few 
membership surveys have extended so far in this direction from the clouds, 
so it is unknown whether these stars are surrounded by a more extensive 
coeval population. Given their proximity in projection to the Perseus 
complex, it seems plausible that they are associated with that more 
distant, but similarly young population. Since both members of each binary 
pair seem to be equally anomalous, this seems to be a reasonable 
explanation for their low positions on the HR diagram; we therefore choose 
to remove them from our sample for all subsequent analysis.

Three of the other binary companions are known sources seen only in 
scattered light due to the presence of an edge-on circumstellar disk. 
Stapelfeldt et al. (1998) used HST and AO observations to show that the 
optical and NIR flux from HK Tau B comes from extended nebulosity, with no 
recognizable flux coming directly from the central star. Krist et al. 
(1995) and Krist et al. (1998) found similar results from HST imaging of 
HL Tau and Haro 6-5B, respectively. HK Tau and Haro 6-5B appear 
significantly underluminous in our HR diagram since this reflected light 
only represents a small fraction of each star's total emitted flux. 
Surprisingly, HL Tau does not appear underluminous, which suggests that 
its luminosity might be dominated by scattered light in the optical and 
direct flux from the central star in the $J$ band. However, we have chosen 
conservatively to omit all three companions from our analysis of relative 
binary ages.

The binary component Haro 6-28 A sits just below the 5 Myr isochrone, so 
further analysis of its scattered-light properties might be worthwhile in 
the future. However, its companion Haro 6-28 B sits just above the 5 Myr 
isochrone, so the inferred ages of the two components are mutually 
consistent. Barring a systematic uncertainty for the binary system, this 
consistency suggests that Haro 6-28 AB is genuinely one of the oldest 
systems in Taurus. Haro 6-28 was originally identified as an H$\alpha$ 
emission line star (Haro et al. 1953; Cohen \& Kuhi 1979) and at least one 
component has a 1.3mm excess indicative of a disk, so its youth seems 
secure.\footnote{As was demonstrated by Carpenter et al. (2006) for Upper 
Sco, massive circumstellar disks are relatively rare ($f\sim$5\%) for 
$\sim$0.3-0.5 $M_{\sun}$ stars by the age of $\sim$5 Myr, though not as 
rare as for higher-mass stars.}

Finally, two companions (V710 Tau C and 2M04554801) sit below the 5 Myr 
isochrone without any obvious explanation. Their optical spectra appear 
to be accurately classified (Luhman 2004; Kraus \& Hillenbrand 2009), so 
a large error in temperature seems unlikely. They are associated with 
stars that appear youthful, so membership in a different population also 
does not explain their anomalously old apparent ages. V710 Tau C has not 
been observed at high spatial resolution, but it shows a very 
significant K band excess in 2MASS ($J-K\sim$2.2), so it might possess 
an edge-on circumstellar disk. 2M04554801 appears to be a point source 
in $K$ band imaging with Keck LGSAO (Kraus \& Hillenbrand, in 
preparation), so if the star is seen in scattered light in the 
near-infrared, then the scattering region must be smaller than for other 
sources with edge-on disks. The original discovery spectrum shows 
obvious signatures of youth (K. Luhman, priv. comm.), and the 
optical/NIR SED does not show any of the characteristic signs of an 
edge-on disk, so the explanation for its underluminosity is currently 
unknown. We can not justify removing either star from our sample, but 
both systems should be regarded with appropriate skepticism.

In the following analysis, we will omit all of the HBC sources that have 
questionable membership. We will retain Haro 6-5B and HL Tau for a more 
in-depth study of high-order multiple systems, but will omit all three 
confirmed scattered light systems for testing coevality.  We will retain 
Haro 6-28 A, V710 Tau C, and 2M04554801 since there is no conclusive 
evidence to suggest that they suffer a systematic bias.

 \begin{figure*}
 \plotone{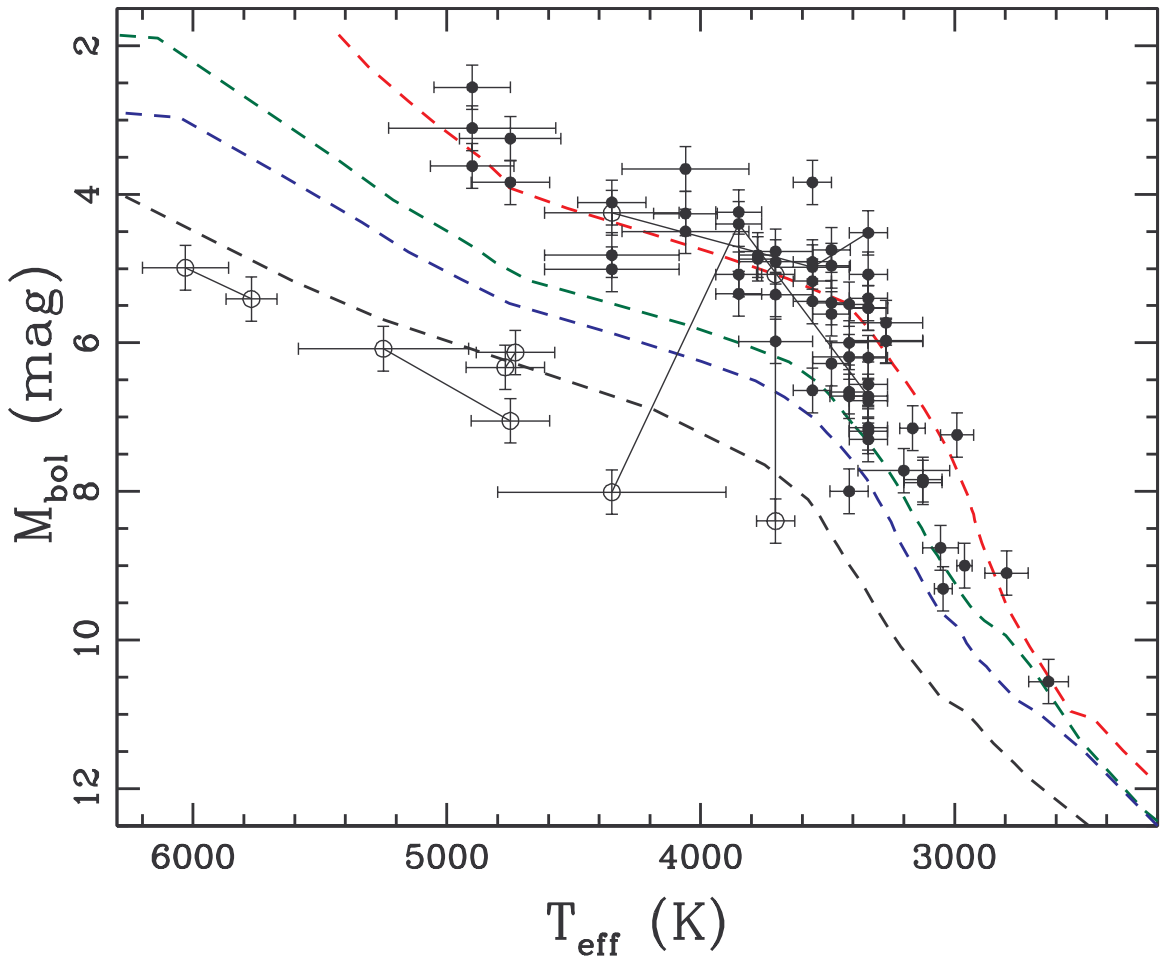}
 \caption{HR diagram for all components of all sample binaries. The binary 
components that we rejected due to known systematic errors (Section 4) are 
shown with open circles, while the rest of our sample is shown with filled 
circles. The dashed lines denote isochrones at 1 Myr (red), 5 Myr (green), 
10 Myr (blue), and 50 Myr (black). Most Taurus members fall along the 1 
Myr isochrone, but 10 fall below the 5 Myr isochrone. Three binary pairs 
that fall mutually below this limit might be associated with the more 
distant Perseus star-forming complex, while individual components that are 
associated with apparently young binary companions could be seen in 
scattered light or have incorrect spectral types. We use solid lines to 
connect each binary pair with one or more rejected components.}
 \end{figure*}

\section{The Coevality of Young Binary Systems}

\subsection{The Relative Ages of Binary Systems}

The first step of our analysis is to determine whether binary systems 
appear more coeval than the association as a whole. The timescale for 
star formation across an entire region could be as long as 10 Myr 
(Mouschovias 1976; Shu 1977), while the formation of a single star 
system should proceed on the dynamical timescale of $\sim$0.1-0.2 Myr (Shu 
et al. 1987). An upper limit on the non-coevality of binary systems will 
provide a direct constraint on the formation timescale for binary 
systems with respect to the entire association.

In Figure 2 (top), we show the same HR diagram as Figure 1, minus the 
objects we eliminated in the discussion above, where each of the binary 
pairs is connected by a line. The overall trend for binary pairs is to 
define lines that roughly parallel the theoretical isochrones, which is 
expected since the models have been chosen by comparison to the Taurus 
single-star sequence. However, as we determine quantitatively below, some 
pairs fall on significantly different isochrones. This type of plot 
provides a summary of the underlying data, but it is hard to draw any firm 
conclusions regarding overall coevality or possible dependence of 
coevality on binary parameters. More detailed statistical analysis must be 
pursued using the inferred stellar ages, as has been concluded by past 
studies of binary ages (e.g Hartigan et al. 1994; White \& Ghez 2001; 
Hartigan \& Kenyon 2003).

In Figure 3 (top), we show a histogram of the absolute difference in the 
inferred logarithmic age, $|\Delta \log \tau| = |\log \tau_{prim} - 
\log \tau_{sec}|$, for each of our binary pairs. The RMS scatter in 
$|\Delta \log\tau|$ among our sample is 0.40 dex and should encompass 
all observational uncertainties as well as any intrinsic age spread for 
binary pairs. Most of the model-predicted masses fall in the range 
$0.3<M<0.9$ $M_{\sun}$, with some masses extending as high as 1.5 
$M_{\sun}$ and as low as 0.03 $M_{\sun}$. We have compared the scatter 
in binary component ages to that of the overall Taurus population by 
using a bootstrap Monte Carlo routine to simulate 10,000 populations 
where we pair each primary with another randomly-selected secondary. In 
Figure 3 (bottom), we show the distribution of all RMS scatter 
measurements for these simulated pairs; only 14 realizations of 
our simulation ($\sim$0.15\%) have RMS scatter of 0.40 dex or less, 
indicating that binaries are more coeval than the overall Taurus 
population at a significance of $\sim$3$\sigma$. The typical age 
differences for binary systems and for random pairs are similar to those 
measured by White \& Ghez (2001); they used a similar sample, but 
estimated effective temperatures using dereddened $V-I$ photometry 
instead of spectral types.

As we describe in Section 3, the observational uncertainties for our 
sample allow us to estimate the corresponding uncertainty in each binary 
component's age, and thus in the degree of coevality. These estimated 
uncertainties vary significantly across our sample, but the median, 
mean, and quadratic mean of these uncertainties (0.33, 0.42, and 0.48 
dex in measured $|\Delta$$\log \tau|$, respectively) are all similar to 
the standard deviation for our sample. This strongly suggests that much 
of the total error budget is dominated by observational errors, and 
therefore that model-related errors and the true dispersion in relative 
ages for binary components are both $<<$0.40 dex.

However, we must also consider whether a single distribution is adequate 
for describing all binary systems, as there are several effects that 
could bias one binary component's age by a significant amount (including 
overluminosity due to unresolved additional multiplicity or 
underluminosity because an object is seen in scattered light). We could 
expect a narrow distribution centered close to zero and broadened by the 
observational errors (corresponding to unaffected binary systems) plus a 
secondary peak away from zero (for systems affected by unusual phenomena 
such as those mentioned above). The observed distribution seems to match 
this expectation, with most systems concentrated at 
$|\Delta$$\log\tau|$$<$0.3 dex and an extended tail at 
$|\Delta$$\log\tau|$$\ga$0.5 dex. If we omit the extended tail and 
compute the standard deviation of the binary age dispersion for only 
systems with $|\Delta$$\log \tau|$$<$0.4 dex, we find a dispersion of 
$\sigma$$_{|\Delta\log\tau|}$$\sim$0.16 dex, corresponding to a typical 
factor of 1.5 in relative age. This dispersion is actually lower than 
the typical uncertainties we estimated above for $|\Delta \log \tau|$, 
which suggests that we might have been too conservative in estimating 
observational uncertainties.

The extended tail includes one sample member that we suggested to be a 
candidate edge-on disk (V710 Tau C) based on its underluminosity and 
extremely red $J-K$ color. The tail also includes two possible hierarchical 
triple systems, DK Tau and XZ Tau, which were suggested to be possible 
hierarchical triples by Jensen et al. (2004) and Carrasco-Gonzalez et al. 
(2009). Polarization measurements by Jensen et al. (2004) indicate that the 
component disks in DK Tau AB are misaligned, unlike most other double-disk 
systems in Taurus. Radio observations by Carrasco-Gonzalez et al. (2009) 
show that XZ Tau B has a double-peaked distribution that could result from a 
$\sim$13 AU binary companion.

In light of this possible bimodality, it is worthwhile to return to Figure 
2 and plot only the ``coeval'' sample ($|\Delta$$\log\tau|$$<$0.4 dex; 
center) and the ``non-coeval'' sample ($|\Delta$$\log\tau|$$>$0.4 dex; 
right). This division reveals a startling trend; among the ``non-coeval'' 
subsample, 11 of the 12 systems have a significantly younger primary star. 
As we show in Appendix A, the model-derived ages in this mass range do not 
show a mass-dependent trend, so the tendency for some binary primaries to 
appear younger must be either a genuine result of the formation process or 
a result of binary-specific systematic uncertainties. Given the clear 
discrepancy with respect to the apparently coeval majority of our sample, 
we strongly suspect that systematic errors are to blame.

The fraction of apparently coeval binary systems in our sample (24/36) 
is identical to the fraction identified by the survey of Hartigan et al. 
(1994). In a sample of binary systems in Taurus and Orion (which 
included many Taurus binaries that we have rejected as hierarchical 
multiples), they found that 17/26 had $\Delta$$\log(L)$$<$0.24 dex, 
corresponding roughly to $|\Delta$$\log\tau|$$\la$0.4 dex. On its face, 
this result suggests that all of the improvements in evolutionary 
models, spectral type assessments, and multiplicity surveys in the past 
15 years have only served to cut the standard deviation in 
$|\Delta$$\log \tau|$ for coeval systems from 0.23 dex to 0.16 dex. 
However, there is one significant difference. All of our non-coeval 
systems possess apparently younger primaries, while all of the 
corresponding systems in the Hartigan et al. sample possess apparently 
older primaries. There is little overlap between our samples since many 
wide Taurus binaries have since been discovered to be hierarchical 
multiples, so one possible explanation is that our stringent 
multiplicity vetting simply allows another systematic error to dominate. 
There are three likely culprits for the systematic error that makes some 
systems appear non-coeval.

First, the non-coevality could result from unrecognized high-order 
multiplicity. The binary fraction is higher among solar-type stars than 
lower-mass stars (Duquennoy \& Mayor 1991; Fischer \& Marcy 1992). If 
this trend also applies to the fragmentation of binary components into 
high-order multiples, then we might expect more binary primaries to be 
unresolved pairs (which would then appear to be a single overluminous 
star). An overluminosity by 0.75 mag (denoting an equal-mass binary 
pair) should correspond to an apparent age discrepancy of $\sim$0.5 dex, 
which matches the observed trend. However, our sample includes many 
objects with similar temperatures (and thus presumably similar masses), 
and as we will show in a future paper (Kraus \& Hillenbrand, in 
preparation), the frequency of binaries at separations $\la$50--100 AU 
in Taurus is nearly constant for all masses $\ga$0.3 $M_{\sun}$. We 
might expect a significant excess of apparently younger primary stars if 
the secondary masses fell significantly below this limit, but few of our 
targets do.

The other two possible explanations are tied to the properties of 
circumstellar disks. Surveys of protoplanetary disks in binary systems 
have suggested that disks might be more likely to form or persist around 
the primary than the secondary (e.g. Monin et al. 2007), even though 
lower-mass stars generally retain their disks longer (e.g. Carpenter et 
al. 2006). We determined most stellar luminosities from the $K$ band flux, 
so a disk excess could have led to significant overestimation of those 
luminosities. In addition, the disks of high-mass stars boast more 
substantial NIR excesses than those of low-mass stars and brown dwarfs 
(e.g. Meyer et al. 1997 versus Liu et al. 2003), so the magnitude of the 
luminosity overestimate should also be larger for primaries than 
secondaries.

All of these explanations should be investigated and ruled out before an 
astrophysical explanation is considered. In particular, systematic effects 
from disks should be mitigated by estimating stellar luminosities using the 
least contaminated filter ($J$) and by modeling the circumstellar dust 
emission using $JHKL$ photometry so that any remaining excess can be 
subtracted. However, our preliminary disk census suggests that NIR excess 
contamination might play only a modest role in biasing relative binary ages 
(unless the disk directly obscures the central star, making it appear in 
scattered light; Section 4). Of the 24 pairs which appear coeval, 20 have at 
least one disk (where $\ge$4 are mixed pairs and $\ge$11 are double-disk 
systems). This presents little contrast to the 12 apparently non-coeval 
pairs, of which nine have at least one disk (with 3 mixed pairs and 6 
double-disk systems). The similar and nontrivial fractions of mixed pairs 
are difficult to explain if disk biases dominate, though double-disk systems 
could appear coeval if both binary components' luminosities are biased 
upward by the same amount.

 \begin{figure}
 \plotone{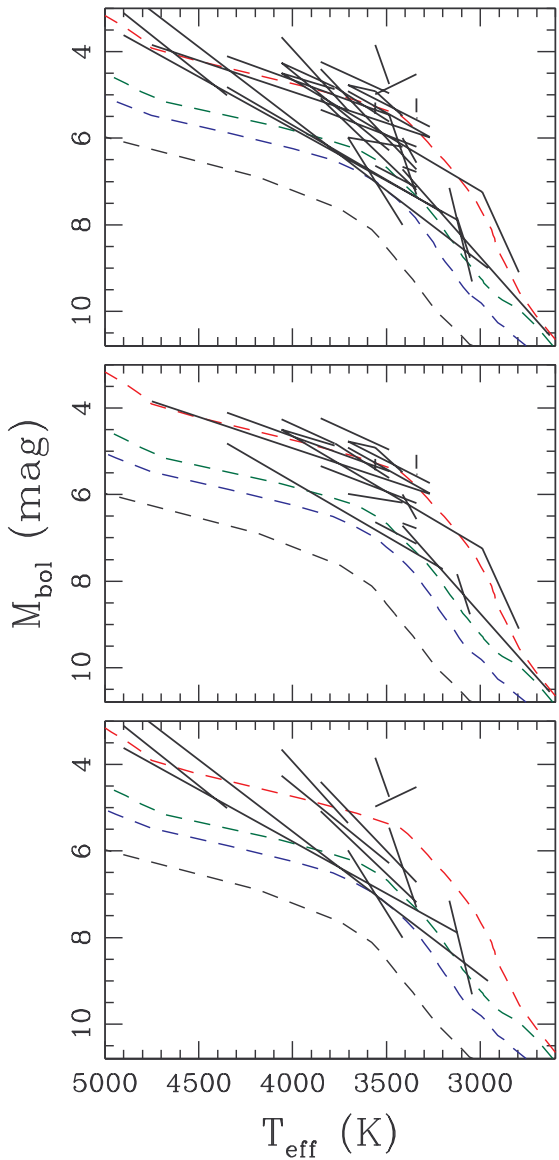}
 \caption{HR diagram for the binary pairs in our sample, where 
each pair is connected by a line. The top panel shows all 
systems, whereas following the text in Section 5.1, the other 
panels show only systems with $|\Delta \log \tau| < 0.4$ dex 
(middle) and $|\Delta \log \tau| > 0.4$ dex (bottom). The binary 
systems in our sample trace the approximate contours of stellar 
evolutionary models, suggesting that the overall trend is correct, 
but our detailed results are more easily described in terms of the 
inferred stellar ages (Figures 3-6).}
 \end{figure}

 \begin{figure*}
 \plotone{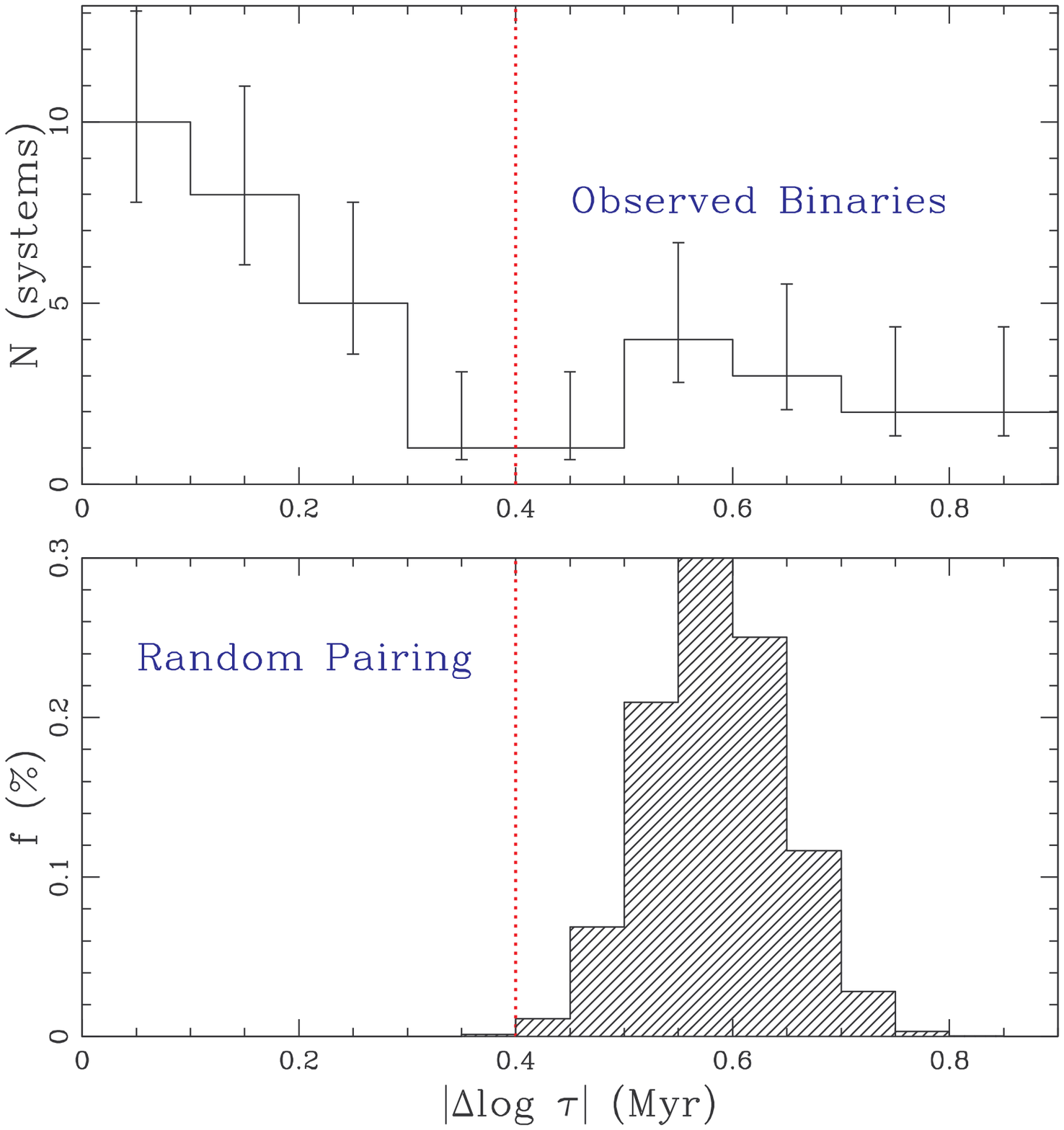}
 \caption{Top: Distribution of differences in logarithmic age, 
$|\Delta \log \tau|$, for all 36 pairs of stars in our sample. The 
RMS scatter in $|\Delta \log \tau|$, $\sigma=0.40$ dex, is indicated 
by a red dotted line. Bottom: Distribution of RMS scatter for a 
set of 10,000 simulated binary populations that were constructed 
by randomly pairing primaries with secondaries. As before, we show 
the RMS scatter of our observed population with a red dotted line; 
only 14 of the 10,000 simulated populations have $\sigma$$\la$0.40 
dex, indicating that our binary pairs are more coeval than Taurus 
at $\sim$3$\sigma$ significance.}
 \end{figure*}

\subsection{The Role of System Parameters in Binary Coevality}

The detailed physics of multiple star formation are still poorly 
understood, so any apparent trends in the coevality of binary systems 
could yield valuable new constraints on theoretical models. A third of 
the systems in our sample appear non-coeval, so any such trend could be 
identified among the 12 non-coeval binary pairs. The three binary 
properties that we can test against system coevality are the component 
mass ratio, the total system mass, and the system projected separation. 
The degree of coevality as a function of separation across the entire 
association could also constrain the large-scale star formation 
processes, so we also analyze the coevality as a function of separation 
between all pairs of stars in our sample.

In Figure 4, we plot the difference in system age $|\Delta$$\log\tau|$ 
as a function of binary mass ratio. If binaries truly formed 
non-coevally, then we might expect the systems with the most extreme or 
most similar mass ratios to show the largest discrepancy in ages. 
However, the dispersion in $|\Delta$$\log\tau|$ for the five systems 
with $q<0.3$ is 0.41 dex, similar to the overall dispersion for our full 
sample (0.40 dex). If we limit this analysis to only the apparently 
coeval population ($|\Delta$$\log\tau|$$<$0.40 dex), the dispersions are 
0.17 dex and 0.16 dex, respectively. This indicates that there is no 
strong trend for reduced coevality in these extreme systems. Hartigan et 
al. (1994) also found no such trend in their sample.

In Figure 5, we plot $|\Delta$$\log\tau|$ as a function of total 
system mass. Most of our sample spans only a limited mass range 
(0.7-1.5 $M_{\sun}$), but we see no evidence of a mass-dependent 
trend. Our sample includes only four systems with a total mass of 
$<$0.5 $M_{\sun}$, but we also see no significant trend for a 
higher scatter in ages. The dispersion (0.36 dex) is similar to 
that of the full sample, though almostly entirely dominated by one 
system (2M04554757+2M04554801). Our sample includes only two 
high-mass pairs, RW Aur AB and V773 Tau Aab, for which we measure 
age discrepancies of 0.55 dex and 0.01 dex, respectively.

In Figure 6, we plot $|\Delta$$\log\tau|$ as a function of system 
separation. If the separation of a binary system scales with the 
protostellar core size when fragmentation occurred, then wider systems 
should typically fragment at an earlier stage than closer systems. This 
suggests that wide pairs might fragment earlier and evolve more 
independently, possibly yielding binary components with a larger 
dispersion in apparent ages. However, as for the previous figures, this 
comparison does not indicate any significant role of separation in 
establishing the binary component ages. The inner and outer halves of the 
sample (divided at 800 AU) have dispersions of 0.37 dex versus 0.42 dex 
(for the full set) and 0.14 versus 0.18 dex (for the coeval subset). We 
conclude that binary systems of all separations are similar coeval to 
within our observed limits.

 \begin{figure}
 \plotone{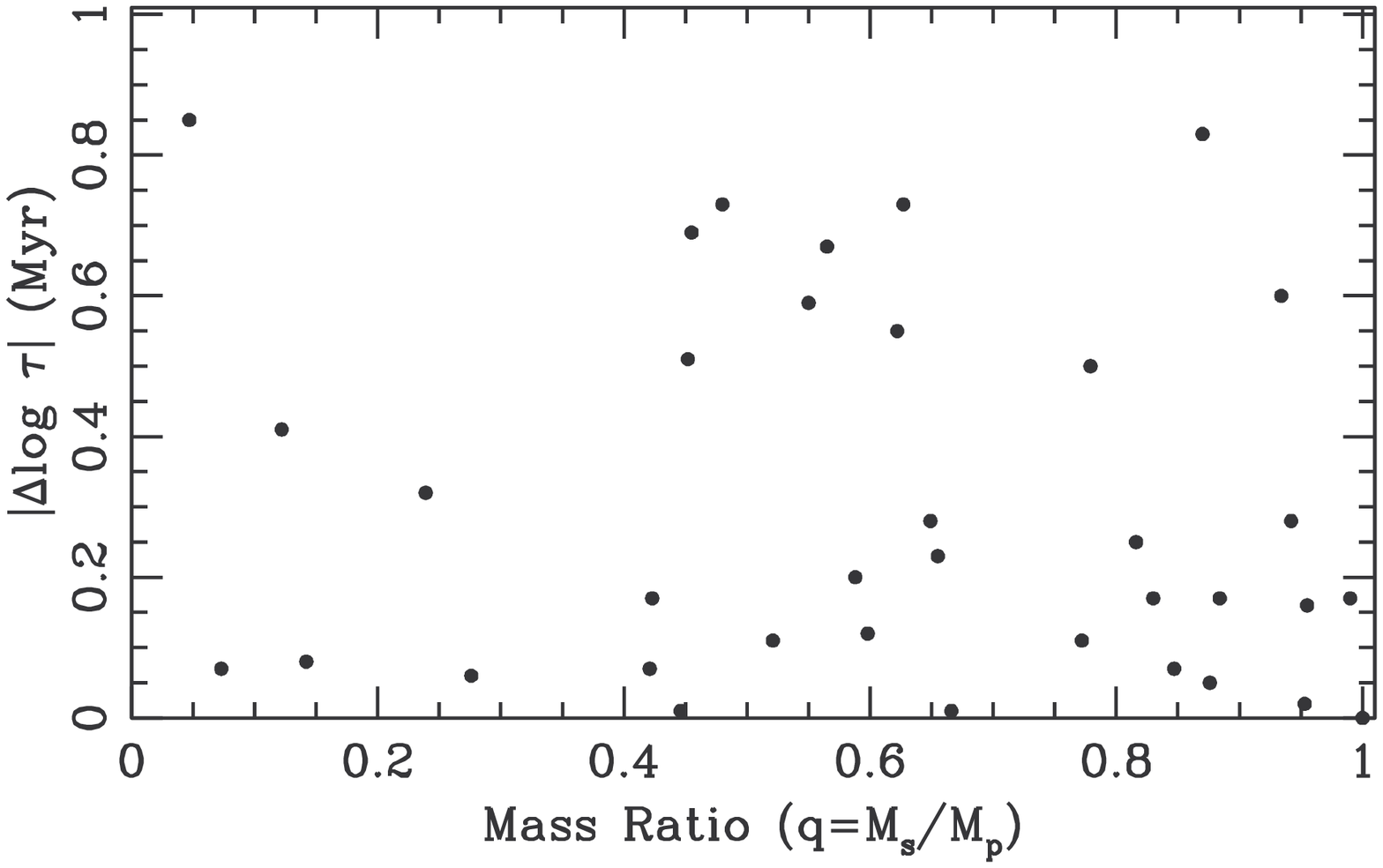}
 \caption{Difference in binary component age, 
$|\Delta$$\log\tau|$, as a function of binary mass ratio. We see 
no evidence of a trend with $q$, as the standard deviation in 
$|\Delta$$\log\tau|$ for $q<0.3$ and for the full sample are the 
same for all pairs (0.41 dex versus 0.40 dex) and for the apparently 
coeval subset (0.17 versus 0.16 dex).}
 \end{figure}

 \begin{figure}
 \plotone{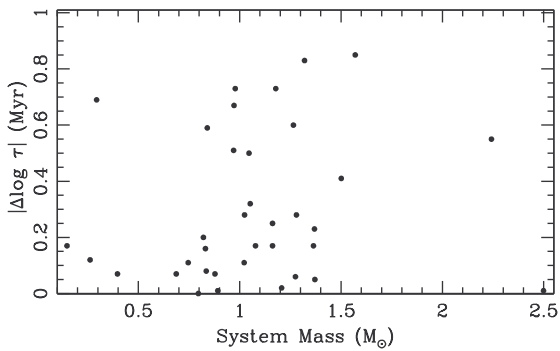}
 \caption{Difference in binary component age, $|\Delta$$\log\tau|$, as a 
function of system mass. We see no 
trend for low-mass systems to appear more discrepant, but are 
unable to test systems with $M\ga$1.5 $M_{\sun}$ and can only 
test a handful of systems with $M\la$0.7 $M_{\sun}$.}
 \end{figure}

 \begin{figure}
 \plotone{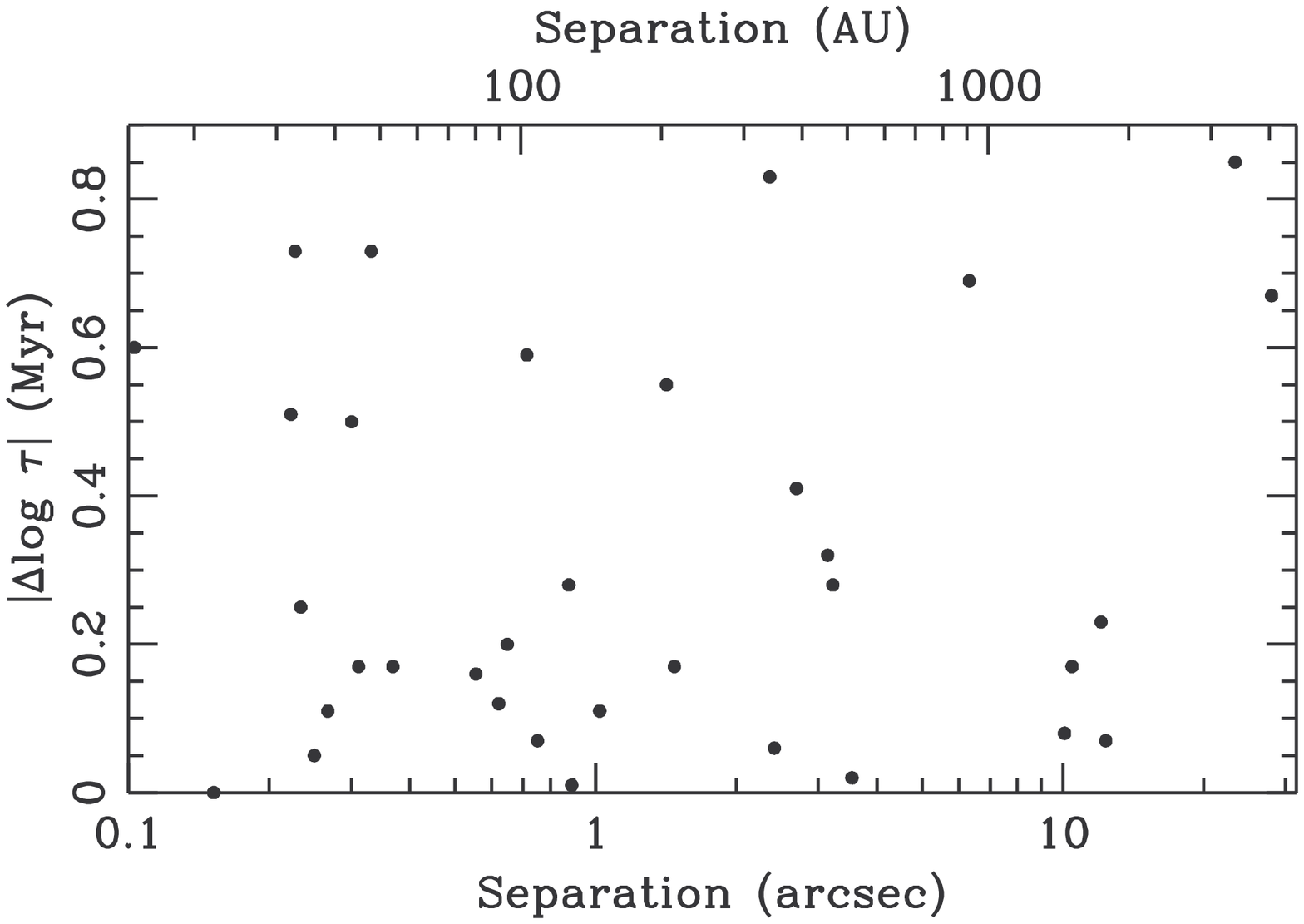}
 \caption{Difference in binary component age, $|\Delta$$\log\tau|$, as a 
function of binary separation. The standard deviations in 
$|\Delta$$\log\tau|$ for the inner and outer halves are 0.37 dex versus 
0.42 dex (for the full set) and 0.14 dex versus 0.18 dex (for the coeval 
subset); in both cases, the inner and outer halves are divided at 800 AU. 
This indicates that binary systems of all separations are similarly 
coeval.}
 \end{figure}

\subsection{The Intra-Association Coevality of Young Stars}

The lack of a separation-dependent trend in differential age begs an 
important question. If binary pairs are similarly coeval with their 
associated components, but significantly more coeval than the association 
as a whole, then what is the form of the transition between these regimes? 
Are adjacent (but unassociated) Taurus members more coeval than distant 
members, or is the age spread similar across all spatial scales greater 
than the binary separation regime? The distribution of Taurus members has 
been suggested to represent a small number of subclusters (e.g. Gomez et 
al. 1993) with radii of $\sim$1-2$^o$ ($\sim$5 pc), though it is unclear 
whether those groups are distinct from the large-scale filamentary (and 
possibly fractal) structure (Kraus \& Hillenbrand 2008). If these apparent 
groupings are closely associated, then stars separated by $<$5 pc might be 
more coeval than the wider association.

In Figure 7, we address this question by plotting $|\Delta$$\log\tau|$ as a 
function of separation for all possible pairs of the primary and secondary 
stars in our binary sample. We can not draw any conclusions for separations 
of 30-1000\arcsec\, (5000 AU to 0.7 pc) due to small number statistics, but 
the dispersion at larger separations (as indicated by the standard deviation 
in bins 0.5 dex wide) is consistently $\sim$0.6 dex across the entire 
separation range. This result suggests that the coevality we see for binary 
systems (Figure 6) is limited to scales of $<$0.7 pc. We can not determine 
if this coevality is limited exclusively to binary systems, though; as we 
showed in our analysis of the spatial distributions of stars (using 
two-point correlation functions; Kraus \& Hillenbrand 2008), the binary 
regime only encompasses separations of $\la$2\arcmin. The observed scatter 
is unlikely to result from any distance dispersion of Taurus members 
($\la$15 pc for members in similar parts of the cloud; Section 3.1) since it 
would yield a scatter of $\la$0.2 mag in $M_{bol}$ or $\sim$0.15 dex in 
$\log\tau$ (0.20-0.25 dex in $|\Delta$$\log\tau|$ for a pair of stars).

 \begin{figure*}
 \plotone{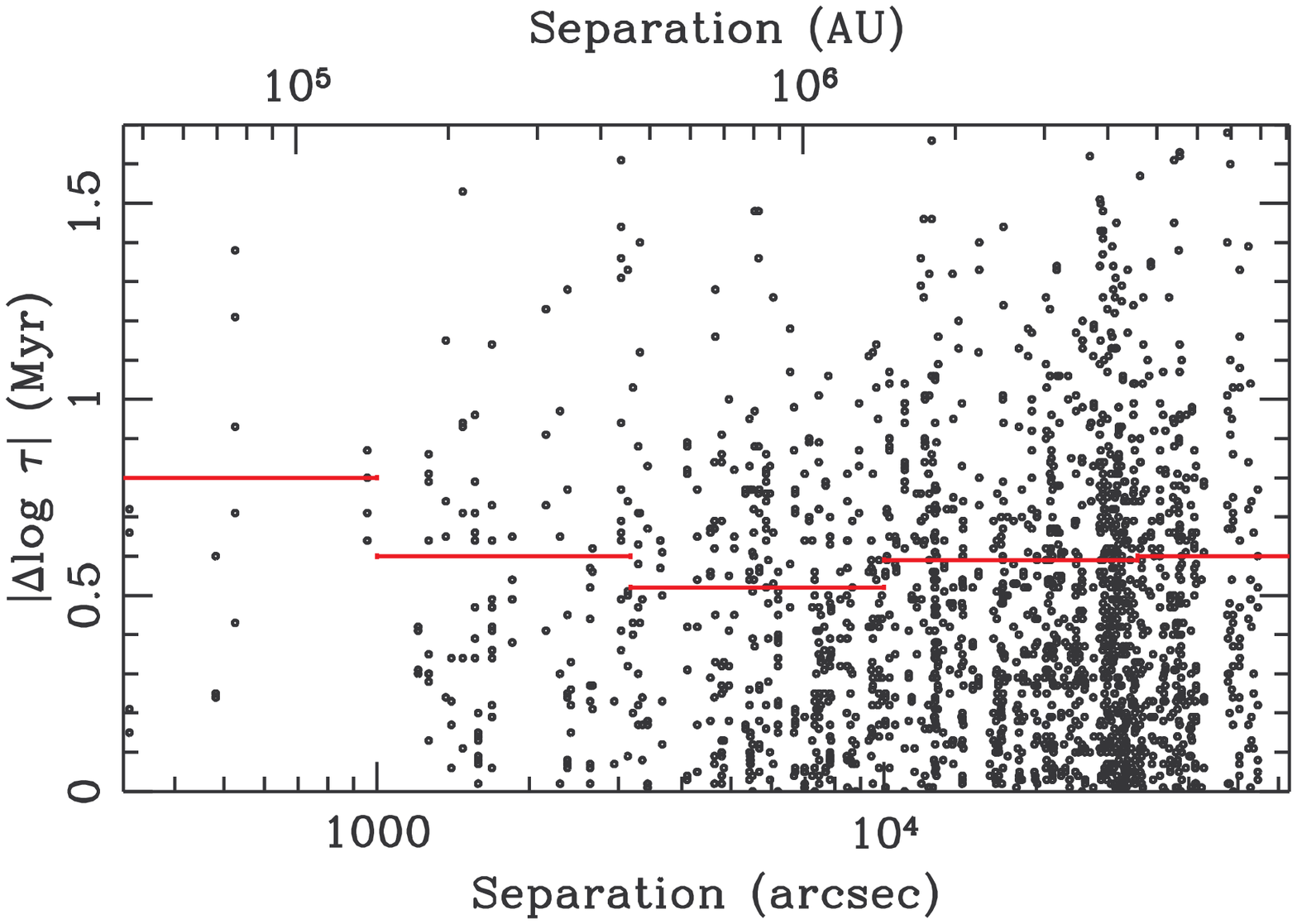}
 \caption{Difference in age as a function of (large-scale) separation 
for all possible pairs of Taurus members among our binary sample. We 
also show the dispersion for all pairs in bins 0.5 dex wide (red lines). 
The sample is insufficient for testing coevality on scales smaller than 
$\sim$1000\arcsec (0.7 pc), but all pairs on larger spatial scales have 
a dispersion of $\sim$0.6 dex, which is similar to the age dispersion 
for random pairs of Taurus members (0.58 dex; Section 5.1). This result 
indicates that the coevality seen for binary systems is limited to 
smaller spatial scales, and perhaps only to binary systems themselves.}
 \end{figure*}

\subsection{Implications for (Multiple) Star Formation}

Our results for binary pair age differences are consistent with 
theoretical predictions for the timescale of local star formation (i.e. 
within one core) and global star formation (spanning the entire molecular 
cloud). If most binary systems in Taurus are coeval to within $<$0.16 dex 
(including observational uncertainties), then given its median age 
(1.8$\pm$0.2 Myr; Appendix A), the formation times for binary components 
must differ by $\la$0.7 Myr. The expected timescale for an individual 
protostar to collapse after achieving supercriticality is the dynamical 
timescale ($\sim$0.1-0.2 Myr; Shu et al. 1987); binary fragmentation is 
likely to occur during this collapse, so our limit is consistent with the 
predicted formation timescale. Recent simulations that exploit 
smoothed-particle hydrodynamic and N-body codes also suggest that most 
binary systems form within $\la$0.5 Myr (Delgado-Donate et al. 2004).

In contrast, the timescale for global star formation is likely to be 
much longer, representing either the turbulent dissipation timescale 
($\sim$1 Myr; Ballesteros-Paredes et al. 1999; Elmegreen 2000) or the 
ambipolar diffusion timescale ($\sim$3-10 Myr; Mouschovias 1976; Shu 
1977). Our limit on the age dispersion of binary pairs is shorter than 
either timescale, while the overall age dispersion for unrelated pairs 
of stars is consistent with the ambipolar diffusion timescale, but only 
marginally with the turbulent dissipation timescale. We found a 
dispersion of $\sim$0.6 dex in $|\Delta$$\log\tau|$ for random pairs, 
corresponding to a dispersion in $\log\tau$ of $\sim$0.4 dex. For the 
median Taurus age of $\sim$1.8 Myr, this corresponds to a typical age 
range of 1-5 Myr.

If our results do reveal two distributions (one population that appears 
coeval and one that does not), then the number of systems in each 
distribution will allow a constraint on the fraction of binary systems 
that are genuinely coeval. Of the 36 pairs of stars that we considered, 
24 are coeval to within $|\Delta$$\log \tau|$$<$0.4 dex, while the other 
12 have ages which are more discrepant. This suggests that 
$\ga$67$^{+7}_{-9}$\% of all binary systems are coeval with a dispersion 
of $\la$0.16 dex. However, many of the non-coeval pairs could be 
affected by systematic errors while being genuinely coeval, so this 
fraction is a lower limit. More intensive study of the apparently 
non-coeval pairs should be a priority; as we discussed above, some stars 
(such as V710 Tau C, DK Tau A, and XZ Tau B) already seem potentially 
affected by systematic errors and might be rejected from our sample 
based on additional followup observations.

Finally, our results indicate that the properties of a binary system 
correlate only modestly with the formation timescale; even extreme 
systems (with very wide separations or disparate masses) appear 
similarly coeval on average. Conversely, unbound pairs of stars that are 
only modestly separated ($\sim$1 pc) show the full age dispersion of the 
association. These trends strongly indicate that binary coevality is a 
natural result of the binary formation process itself, not a reflection 
of any trend for star formation to occur simultaneously within larger 
regions of the natal molecular cloud.

\section{Summary}

We have studied the binary population of the Taurus-Auriga association in 
order to quantify the frequency and degree of noncoevality in young binary 
systems. After identifying and rejecting the systems that are known to be 
affected by systematic errors (such as further multiplicity or obscuration 
by circumstellar material), we used pre-main sequence evolutionary tracks 
to infer individual ages for the individual stars in binary systems, and 
hence the relative binary ages. We have found that the overall dispersion 
in the relative ages ($|\Delta$$\log\tau|$) is 0.40 dex, though the 
distribution actually appears bimodal. Random pairs of Taurus members are 
coeval only to within 0.58 dex, suggesting that Taurus binaries are more 
coeval than the association as a whole.

The bimodality indicates that our sample is comprised of two populations, 
with $\sim$2/3 appearing to be coeval binaries with a dispersion of 
$\sigma$$_{|\Delta\log\tau|}$$\sim$0.16 dex and the other $\sim$1/3 
appearing to be systematically offset from coevality by $\sim$0.6 dex. The 
non-coeval population shows no trends with respect to the system mass, 
separation, or mass ratio, which defies the predictions of many formation 
scenarios for non-coeval systems. We therefore suggest that the non-coeval 
population is comprised mainly of unrecognized hierarchical multiples, 
stars seen in scattered light, or stars with NIR disk excesses; 
identifying any truly non-coeval systems will require additional followup 
to rule out or correct for these explanations. The full range of apparent 
ages in our sample is $\sim$1 dex, which suggests that a binary system 
tends to form in a very short period of time relative to the global star 
formation timescale for Taurus.

Finally, we found that the relative coevality of binary systems does not 
depend significantly on the system mass, mass ratio, or separation. 
However, any pair of Taurus members wider than $\sim$10\arcmin\, 
($\sim$0.7 pc) shows the full age spread of the association. This suggests 
that the enhanced coevality is seen only for binary systems and not for 
neighboring stars that formed from separate protostellar cores. The 
apparent coevality of a large fraction of our sample is also a partial 
endorsement of pre-main sequence isochrones. We did invoke several 
corrections to the model tracks, but any additional mass-dependent error 
in ages would cause systems with unequal mass ratios to appear less 
coeval, and we see no such trend to within the uncertainties in our 
results.

\acknowledgements

The authors thank R. White and G. Herczeg for helpful feedback on several 
of the ideas presented here, as well as E. Mamajek for a helpful 
discussion of Taurus membership issues. We also thank the referee for a 
prompt and helpful review. ALK was supported by a NASA Origins grant to 
LAH and by a SIM Science Study. This work makes use of data products from 
2MASS, which is a joint project of the University of Massachusetts and the 
IPAC/Caltech, funded by NASA and the NSF. This work also made extensive 
use of the SIMBAD database, operated at CDS, Strasbourg, France.

\begin{appendix}

\section{The Single Stars in Taurus}

The single stars of Taurus provide a useful check on the validity of our 
results, as well as providing their own constraints on its star formation 
history. In this appendix, we compile a sample of all stars which have a 
significant probability of being single (based on nondetections with one 
or more high-resolution imaging techniques). We then place these single 
stars on an HR diagram and estimate their ages and masses. Finally, we 
investigate the dependence of apparent age on stellar mass and on location 
within the association.

\subsection{Sample}

We list our sample of apparently single stars in Table 3, including 
all of the references for our adopted parameters (singleness, spectral 
type, and extinction). We based our sample on the compilation of all 
Taurus members that we originally described in Kraus \& Hillenbrand 
2007a), and then we omitted all stars that did not have at least one 
observation at high angular resolution. We then searched the 
literature for spectral types, requiring uncertainties of $\le$1 
subclass for spectral types $>$K0 and $\le$2 subclasses for 
earlier-type stars.

Most stars have only been surveyed for multiplicity to a separation limit 
of 50-100 mas (7-15 AU), so close binary systems still contaminate this 
sample. However, we do not expect any mass-dependent systematic biases. 
The binary fraction drops significantly with declining primary mass (e.g. 
Kraus et al. 2006, 2008), but much of that drop is seen among the wider 
binary systems (e.g. Kraus \& Hillenbrand 2009) that would have fallen 
outside our required sensitivity limit of $\la$15-20 AU.

We inferred the fundamental properties of these stars (luminosity and 
temperature, then age and mass) using the methods described in Section 3. 
We list these properties in Table 3.

\subsection{The Ages of Single Taurus Members}

In Figure A1, we show the HR diagram for our sample of single Taurus 
members. The composition of our single-star sample is significantly 
different from our binary sample, featuring many high-mass ($>$1 
$M_{\sun}$) and low-mass ($<$0.3 $M_{\sun}$) members, but few members with 
intermediate masses. This difference in composition is driven largely by 
selection biases since few binaries at either extremum of mass have 
spatially resolved spectra. High-mass binaries in Taurus tend to be 
hierarchical multiples with additional components, while low-mass binaries 
were difficult to observe with spectroscopy before the recent advent of 
laser guide star AO. As a result, we must be very cautious in comparing 
the bulk properties of both samples.

We also note that many members fall below our designated lower edge of the 
Taurus sequence (the 5 Myr isochrone) and might have erroneously low 
luminosities. Most of these members have not been well-studied (e.g. 
ITG33a and I04301+2608), so we can not reject them with certainty, but 
their presence as extreme outliers in our plots invites skepticism. Most 
of the highest-mass members (SpT$\le$K3; $M\ga$1 $M_{\sun}$) also fall 
systematically below the 1-2 Myr isochrone. Few of our binary sample 
members fall in this mass range, but those that do are either obviously 
erroneous (the HBC stars) or appear genuinely young. We have no 
satisfactory explanation for this discrepancy between the single stars and 
binary components since an error in our methods or in the underlying 
models should affect both populations equally, but the small number of 
high-mass stars in our binary sample suggests that the single-star sample 
might provide a more reliable indication of the true empirical isochrone.

In Figure A2, we show the model-derived age as a function of model-derived 
mass for our sample of single stars. Most stars seem to track the median 
age of Taurus, but as we noted above, the highest-mass stars ($\ga$1 
$M_{\sun}$) appear systematically older. The brown dwarfs of Taurus 
($M\la$0.07 $M_{\sun}$) have very uncertain ages, so it is difficult to 
determine when they formed in relation to the stars. This uncertainty is 
driven by the physics of brown dwarf contraction, as isochrones at ages of 
$\la$10 Myr follow similar tracks in the HR diagram. If we only consider 
members with masses of $\sim$0.07-0.9 $M_{\sun}$, then the median age of 
Taurus is $\log(\tau)$$=6.25\pm0.05$ yr ($1.8\pm0.2$ Myr).

Finally, in Figure A3, we show the spatial distribution of our single-star 
sample on the sky, with the position of each star color-coded according to 
its age, as well as the mean age for the eastern subgroup, southern 
subgroup, and the eastern and western halves of the central filaments. The 
eastern subgroup appears $\sim$4$\sigma$ older than the other subgroups 
(3.2 Myr, versus 1.4-1.9 Myr); if this age difference is genuine, then it 
suggests that star formation occurred first in Auriga, then in the rest of 
the association. However, this apparent age difference could also be a 
three-dimensional projection effect; the difference of $\sim$0.25 dex in 
mean age could be explained if the distance to Auriga stars were 
$\sim$15-20\% larger than the mean distance to Taurus ($\sim$170 pc versus 
$\sim$145 pc). Otherwise, there is no significant trend in the ages of 
Taurus members, suggesting that global star formation proceeded nearly 
simultaneously (to within $\la$0.3 Myr).

 \begin{figure*}
 \figurenum{A1}
 \plotone{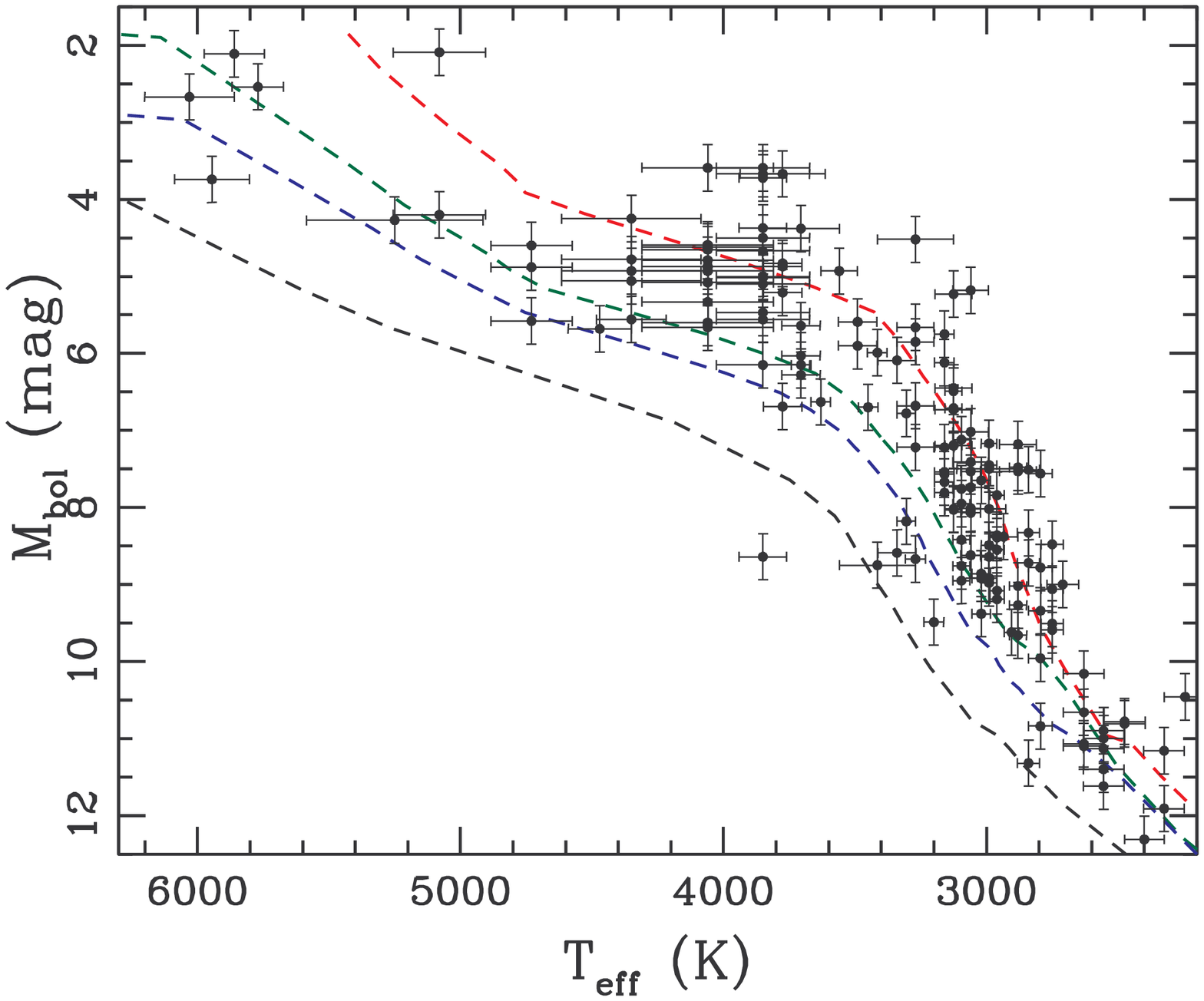}
 \caption{HR diagram for all members of our single star sample. The 
dashed lines denote isochrones at 1 Myr (red), 5 Myr (green), 10 Myr 
(blue), and 50 Myr (black). Most Taurus members fall along the 1-2 Myr 
isochrone, but many fall significantly below that level, perhaps due 
to the presence of an edge-on disk, undiscovered binary companion, 
erroneous observations. The highest-mass stars ($\ga$1 $M_{\sun}$) 
also fall systematically below the 1-2 Myr isochrone, suggesting 
either that the models might not be calibrated correctly in this 
regime or that these stars formed earlier in Taurus.}
 \end{figure*}

 \begin{figure*}
 \figurenum{A2}
 \plotone{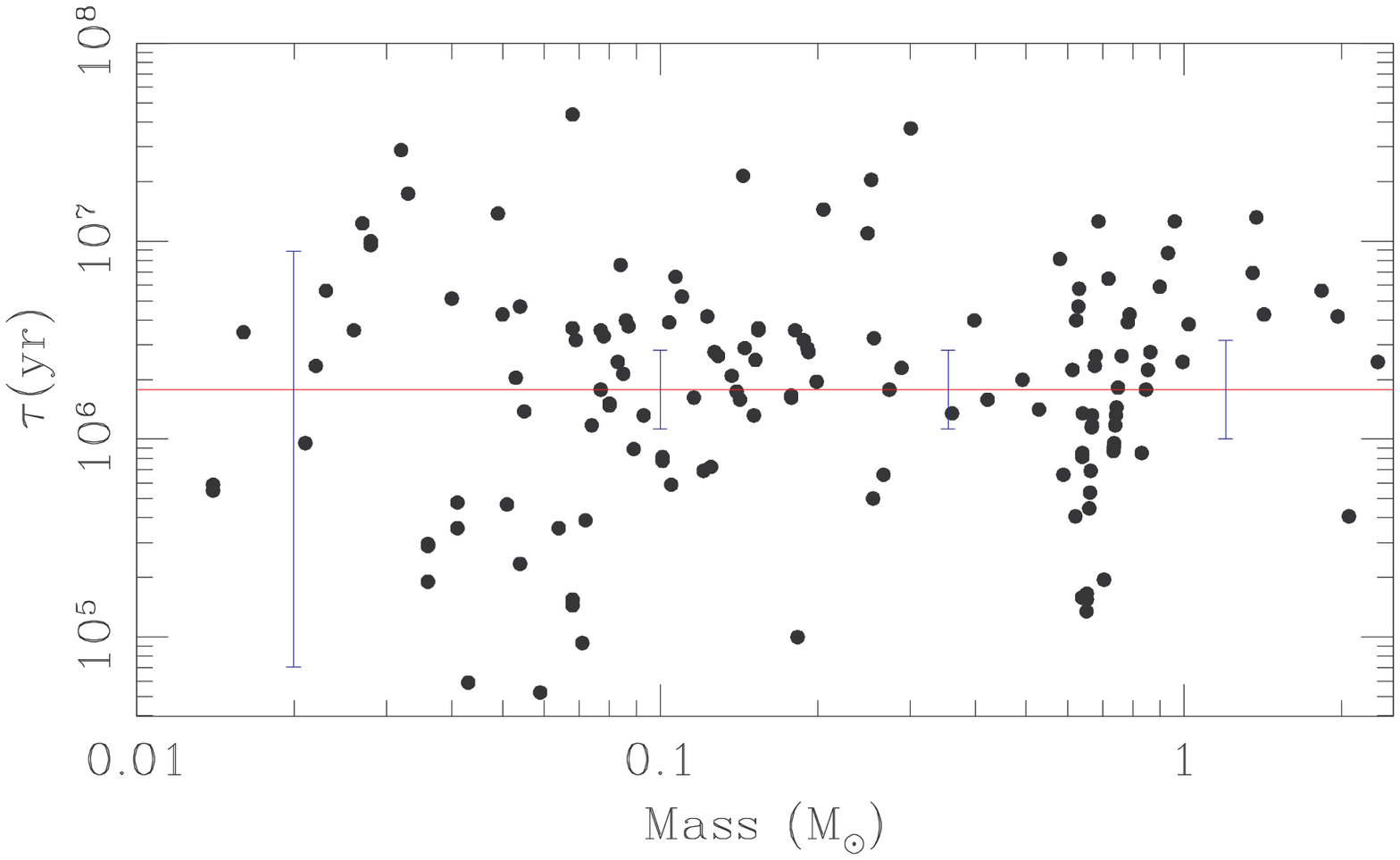}
 \caption{Age as a function of mass for all members of our single star 
sample. We also show the model-derived median age of Taurus (1.8 Myr; red 
line) as determined from our sample, plus representative error bars at 
four different masses (blue). The mass-dependent age of our sample tracks 
the overall median age except at the high-mass end ($\ga$1 $M_{\sun}$), 
where stars appear older, and at the low-mass end ($\la$0.07 $M_{\sun}$), 
where the uncertainties become very large.}
 \end{figure*}

 \begin{figure*}
 \figurenum{A3}
 \plotone{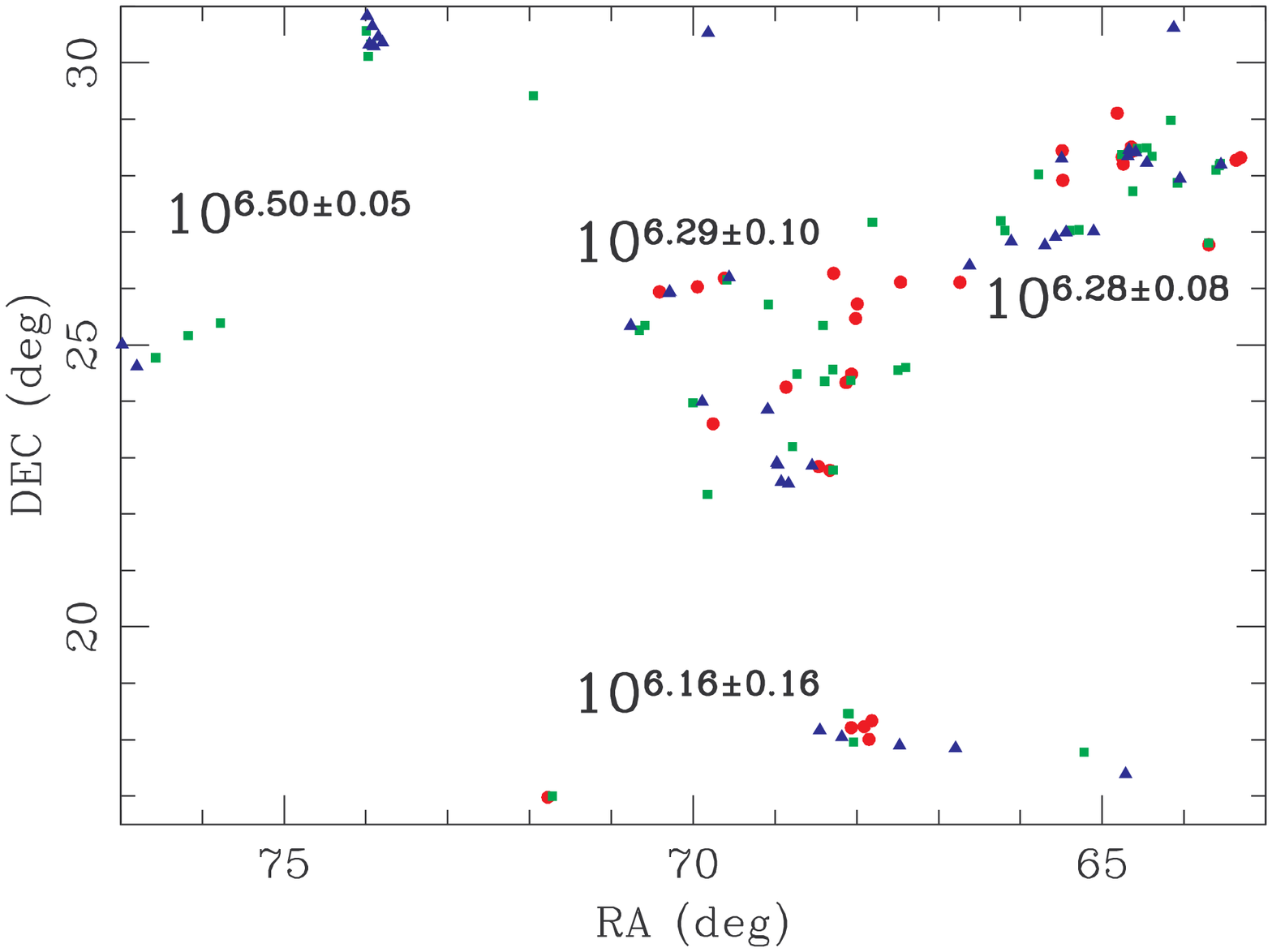}
 \caption{Spatial distribution for all members of our single star 
sample in the mass range that is well-calibrated (0.07-0.9 
$M_{\sun}$), color-coded by age (red circles: $<$1 Myr; green squares: 
1-3 Myr; blue triangles: $>$3 Myr). We also show the mean age for the 
eastern subgroup, southern subgroup, and the eastern and western 
halves of the core regions. The eastern subgroup appears 
$\sim$4$\sigma$ older than the other subgroups (3.2 Myr, versus 
1.4-1.9 Myr) and contains no stars with an apparent age of $<$1 Myr, 
but otherwise there is no apparent pattern in the ages of Taurus 
members.}
 \end{figure*}

\section{The Coevality of Triple and Quadruple Systems}

High-order multiple systems are a critical tool for constraining stellar 
evolutionary models. If these multiple systems form coevally, then they 
provide a simultaneous test of the models at three or more masses. This 
feature was exploited by White et al. (1999) to constrain models with the 
well-known quadruple system GG Tau and to infer the best set of models to 
use for low-mass stars (the Lyon models) as well as to establish the best 
temperature scale for young stars (Luhman et al. 2003). We now extend this 
analysis to a quadruple system (UZ Tau), a quadruple that is part of a 
sextuple system (V955 Tau + LkHa332/G2), three components each of two 
systems that are not yet completely characterized (FV Tau and V773 Tau), 
and three triple systems (FS Tau, V710 Tau, and HL Tau/XZ Tau), plus we 
replicate the analysis of White et al. (1999) for GG Tau to provide 
context.

As we show in Figure B1, all three of the quadruple systems appear to have 
consistent ages. The consistency of GG Tau is partly a result of its 
previous role in calibrating stellar models and temperature scales, but UZ 
Tau appears to be almost coeval and similarly consistent. Three components 
of V955 Tau + LkHa332/G2 also fall along the 1 Myr isochrone, but V955 Tau 
B has an inferred age of $\sim$3 Myr. This is $\sim$2$\sigma$ away from a 
consistent age, but among 12 components, we would expect $\sim$0.6 
outliers at $\ga$2$\sigma$.

Dynamical masses are available in the literature for the UZ Tau Aab and GG 
Tau Aab pairs and can be compared to those inferred from the HR diagram. 
The consistency is mixed. Guilloteau et al. (1999) found from the 
circumbinary disk kinematics that the total system mass for GG Tau Aab is 
1.28$\pm$0.07 $M_{\sun}$; the total mass predicted by theoretical models 
(1.37 $M_{\sun}$) agrees to within $\sim$7\%. By contrast, Prato et al. 
(2002) reported dynamical masses for UZ Tau Aa and Ab of 1.02$\pm$0.06 
$M_{\sun}$ and 0.29$\pm$0.03 $M_{\sun}$, while the masses predicted by 
theoretical models are 0.61 and 0.30 $M_{\sun}$. The secondary mass agrees 
very well, but the discrepancy in the primary mass is very puzzling 
because its position in the HR diagram is virtually identical to that of 
GG Tau Ab, which has excellent consistency between observations and 
theory. Prato et al. explored the possible sources of this discrepancy in 
much greater detail, so we simply note its existence as proof that HR 
diagram analysis plays a critical, but incomplete role in constraining 
stellar evolutionary models. A full study of evolutionary models must 
include their dynamical masses (e.g. Schaefer et al. 2008), not just their 
temperatures and luminosities. Truly precise tests will also require 
direct measurement of radii (Stassun et al. 2008) rather than indirect 
estimates from the Stefan-Boltzmann law and the observed luminosity and 
temperature.

None of the three-component tests in our sample provide the same 
consistency seen among the quadruple systems, though as we described in 
Section 3, the two largest discrepancies are likely to be systematic. The 
edge-on disk host Haro 6-5B sits very far below the Taurus sequence, 
unlike FS Tau AB, while V710 Tau C might also be seen in scattered light. 
HL Tau is also seen in scattered light in the optical, but our inferred 
age based on its $J$ magnitude seems consistent with that of XZ Tau A, 
suggesting that the central star of HL Tau might dominate its luminosity 
in the NIR. As we described in Section 2, XZ Tau B was suggested to be a 
possible binary pair by Carrasco-Gonzalez et al. (2009), which would 
explain its apparent overluminosity. FS Tau A and FS Tau B also appear 
moderately discrepant, sitting 1.5$\sigma$ on either side of the 1 Myr 
isochrone. FV Tau Aa and FV Tau Ba have very consistent ages, but FV Tau 
Bb sits somewhat lower in the HR diagram; the only spatially resolved 
spectrum for FV Tau Bb is very noisy, so the apparent underluminosity 
could actually indicate that it has a later spectral type (M5.0-M5.5 
rather than M3.5).

 \begin{figure*}
 \figurenum{B1}
 \plotone{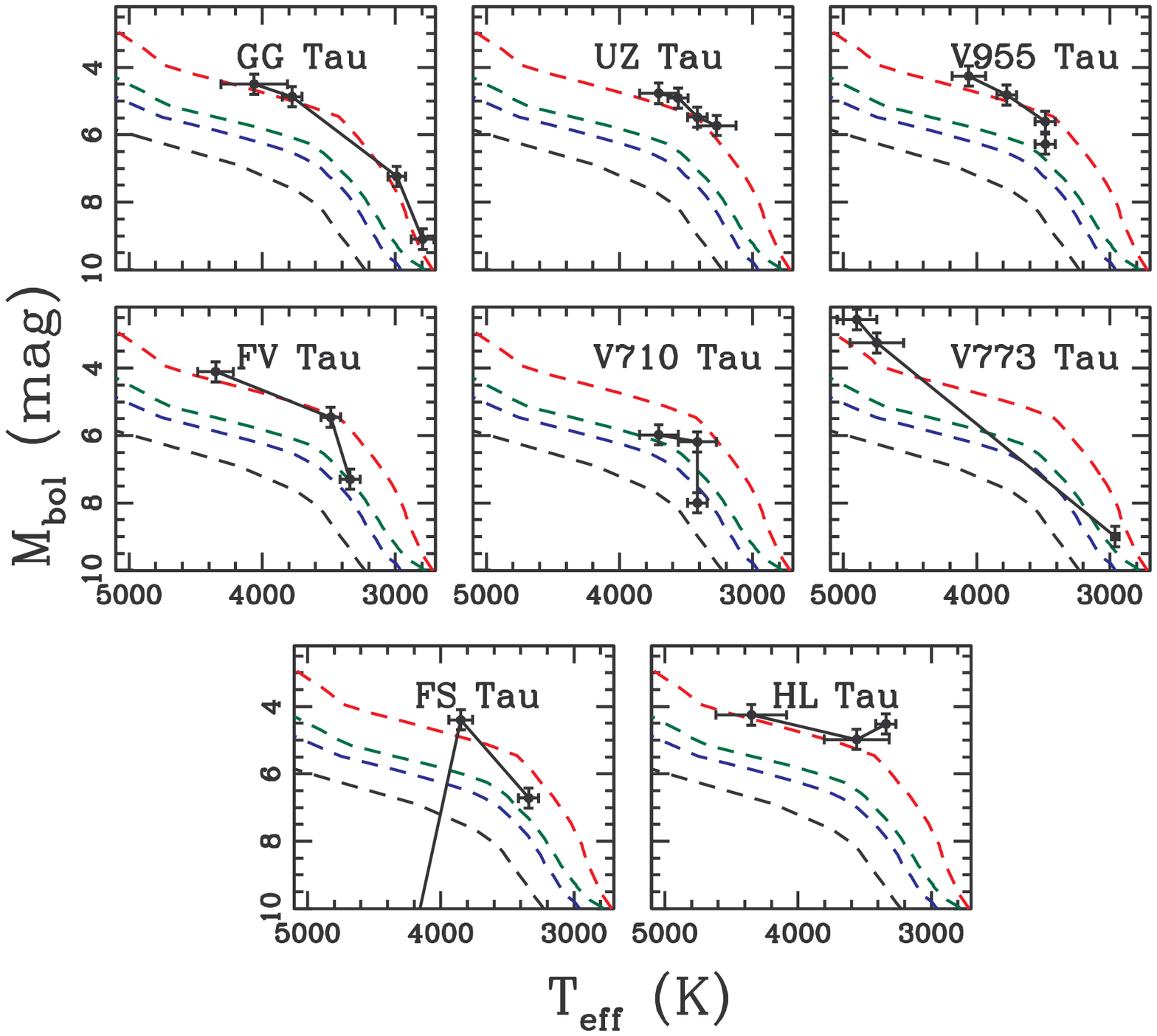}
 \caption{HR diagrams showing the components of eight hierarchical 
multiple systems. The four components of GG Tau and UZ Tau appear to be 
coeval, plus the components of V955 Tau might be coeval. However, as we 
describe in the text, the other five systems all have one or more 
components that disagree significantly. This could be due to errors in 
determining their luminosity (stars seen in scattered light only or which 
host a circumstellar disk) or temperature (incorrect spectral types).}
 \end{figure*}

\end{appendix}

\clearpage

\begin{LongTables}
\begin{landscape}

\begin{deluxetable*}{lllcllrlrcl}
\tablewidth{0pt}
\tablecaption{Binary Sample: Observed and Inferred Properties}
\tabletypesize{\scriptsize}
\tablehead{
\colhead{Name} & \colhead{RA} & \colhead{Dec} & \colhead{Sep} & 
\colhead{Flux} & \colhead{SpT} & \colhead{$A_V$} & 
\colhead{$T_{eff}$} & \colhead{$M_{bol}$} & 
\colhead{Warm} & \colhead{Refs}
\\
\colhead{} & \multicolumn{2}{c}{(J2000)} & \colhead{(mas)} & 
\colhead{(mag)} & \colhead{} & \colhead{(mag)} & \colhead{(K)} & \colhead{(mag)} & \colhead{Disk?}
}
\startdata
HBC 352&3 54 29.51&+32 03 01.4&8970$\pm$70&J=10.09&G0$\pm$2&0.9&6030$\pm$170&4.99&...&1, 2\\
HBC 353&3 54 30.17&+32 03 04.3&8970$\pm$70&J=10.45&G5$\pm$2&1.0&5770$\pm$100&5.41&...&1, 2\\
HBC 355&3 54 35.97&+25 37 08.1&6310$\pm$70&J=10.81&K0$\pm$2&0.5&5250$\pm$335&6.08&...&1, 3\\
HBC 354&3 54 35.56&+25 37 11.1&6310$\pm$70&J=11.80&K3$\pm$1&1.2&4750$\pm$155&7.05&...&1, 3\\
HBC 356&4 03 13.96&+25 52 59.8&1280$\pm$20&J=10.84&K3$\pm$1&0.7&4750$\pm$155&6.23&N&4, 5, 20\\
HBC 357&4 03 13.96&+25 52 59.8&1280$\pm$20&J=10.84&K3$\pm$1&0.7&4750$\pm$155&6.23&N&4, 5, 20\\
V773 Tau Aa&4 14 12.92&+28 12 12.4&SB&K=6.72&K2$\pm$1&1.8&4900$\pm$150&2.56&Y?&6, 20\\
V773 Tau Ab&4 14 12.92&+28 12 12.4&SB&K=7.27&K3$\pm$1.5&1.8&4750$\pm$200&3.25&Y?&6, 20\\
2M04141188&4 14 11.88&+28 11 53.5&23380$\pm$70&J=13.16&M6.25$\pm$0.25&1.0&2960$\pm$30&9.00&...&3, 7\\
FO Tau A&4 14 49.29&+28 12 30.6&152.5$\pm$2.9&K=8.87&M3.5$\pm$0.5&1.9&3340$\pm$75&5.53&Y&8, 9\\
FO Tau B&4 14 49.29&+28 12 30.6&152.5$\pm$2.9&K=8.87&M3.5$\pm$0.5&1.9&3340$\pm$75&5.53&Y&8, 9\\
DD Tau A&4 18 31.13&+28 16 29.0&555$\pm$10&K=8.45&M3.5$\pm$0.5&2.1&3340$\pm$75&5.08&Y&8, 9, 21\\
DD Tau B&4 18 31.13&+28 16 29.0&555$\pm$10&K=8.85&M3.5$\pm$0.5&2.9&3340$\pm$75&5.40&Y&8, 9, 21\\
FQ Tau A&4 19 12.81&+28 29 33.1&752$\pm$14&K=10.03&M3$\pm$0.5&2.0&3415$\pm$75&6.66&Y&8, 9, 21\\
FQ Tau B&4 19 12.81&+28 29 33.1&752$\pm$14&K=10.11&M3.5$\pm$0.5&1.8&3340$\pm$75&6.78&Y&8, 9, 21\\
LkCa 7 A&4 19 41.27&+27 49 48.5&1021$\pm$19&K=8.74&M0$\pm$0.5&0.2&3850$\pm$90&5.34&N&8, 9, 20\\
LkCa 7 B&4 19 41.27&+27 49 48.5&1021$\pm$19&K=9.37&M3.5$\pm$0.5&0.4&3340$\pm$75&6.20&N&8, 9, 20\\
FS Tau A&4 22 02.18&+26 57 30.5&227.6$\pm$7.1&K=8.33&M0$\pm$0.5&5.0&3850$\pm$90&4.40&Y&8, 9\\
FS Tau B&4 22 02.18&+26 57 30.5&227.6$\pm$7.1&K=10.43&M3.5$\pm$0.5&5.2&3340$\pm$75&6.72&Y&8, 9\\
Haro 6-5B&4 22 00.69&+26 57 33.3&19880$\pm$70&J=15.08&K5$\pm$2&10.0&4350$\pm$450&8.01&Y&10, 11\\
FV Tau A&4 26 53.53&+26 06 54.4&12081$\pm$9&J=9.92&K5$\pm$0.5&5.4&4350$\pm$135&4.11&Y&8, 9, 21\\
FV Tau/c A&4 26 54.41&+26 06 51.0&12081$\pm$9&K=9.00&M2.5$\pm$0.5&3.3&3485$\pm$75&5.46&N&8, 9, 21\\
FV Tau/c B&4 26 54.41&+26 06 51.0&713$\pm$1.8&K=11.21&M3.5$\pm$0.5&7.0&3340$\pm$75&7.30&Y&8, 9, 21\\
DF Tau A&4 27 02.80&+25 42 22.3&103$\pm$2&K=7.13&M2$\pm$0.5&0.6&3560$\pm$75&3.84&Y&5, 8\\
DF Tau B&4 27 02.80&+25 42 22.3&103$\pm$2&K=8.01&M2.5$\pm$0.5&0.8&3485$\pm$75&4.75&Y&5, 8\\
2M04284263 A&4 28 42.63&+27 14 03.9&621$\pm$7&K=10.85&M5$\pm$0.5&0.5&3125$\pm$75&7.84&Y?&7, 12, 23\\
2M04284263 B&4 28 42.63&+27 14 03.9&621$\pm$7&K=11.75&M5.5$\pm$0.5&0.5&3055$\pm$70&8.76&Y?&3, 7, 12, 23\\
UX Tau A&4 30 04.00&+18 13 49.4&5856$\pm$3&K=7.60&K2$\pm$1&0.2&4900$\pm$165&3.62&Y&1, 13, 21\\
UX Tau C&4 30 04.00&+18 13 49.4&2692$\pm$2&K=10.85&M5$\pm$0.5&0.1&3125$\pm$75&7.88&N&1, 13, 14, 21\\
FX Tau A&4 30 29.61&+24 26 45.0&890$\pm$17&K=8.33&M1$\pm$1&1.1&3705$\pm$145&4.91&Y&4, 9, 21\\
FX Tau B&4 30 29.61&+24 26 45.0&890$\pm$17&K=9.19&M4$\pm$1&1.1&3270$\pm$145&5.97&N&4, 9, 21\\
DK Tau A&4 30 44.25&+26 01 24.5&2360$\pm$1&K=7.36&K9$\pm$1&0.8&4060$\pm$250&3.66&Y?&4, 13, 22\\
DK Tau B&4 30 44.25&+26 01 24.5&2360$\pm$1&K=8.74&M1$\pm$1&0.8&3705$\pm$145&5.35&Y?&4, 13, 22\\
V927 Tau A&4 31 23.82&+24 10 52.9&267$\pm$6.8&K=9.31&M3$\pm$0.5&1.4&3415$\pm$75&6.00&N&8, 9, 23\\
V927 Tau B&4 31 23.82&+24 10 52.9&267$\pm$6.8&K=9.79&M3.5$\pm$0.5&0.9&3340$\pm$75&6.56&N&8, 9, 23\\
HL Tau&4 31 38.44&+18 13 57.7&23310$\pm$70&J=10.62&K5$\pm$1&7.4&4350$\pm$265&4.25&Y&10, 15\\
XZ Tau A&4 31 40.07&+18 13 57.2&300.6$\pm$1.3&K=8.36&M2$\pm$1&1.4&3560$\pm$145&4.98&Y&8, 9\\
XZ Tau B&4 31 40.07&+18 13 57.2&300.6$\pm$1.3&K=7.80&M3.5$\pm$0.5&1.4&3340$\pm$75&4.52&Y&8, 9\\
HK Tau A&4 31 50.57&+24 24 18.1&2342$\pm$61&K=8.64&M1$\pm$0.5&2.3&3705$\pm$75&5.08&Y&1, 9, 10, 23\\
HK Tau B&4 31 50.57&+24 24 18.1&2342$\pm$61&K=11.96&M1$\pm$0.5&2.3&3705$\pm$75&8.40&Y&1, 9, 10\\
V710 Tau A&4 31 57.79&+18 21 38.1&3224$\pm$3&K=9.38&M1$\pm$1&0.9&3705$\pm$145&5.98&Y&1, 13, 21\\
V710 Tau B&4 31 57.79&+18 21 38.1&3224$\pm$3&K=9.44&M3$\pm$1&0.9&3415$\pm$145&6.19&N&1, 13, 21\\
V710 Tau C&4 31 57.79&+18 21 38.1&28000$\pm$70&J=12.26&M3$\pm$0.5&0.9&3415$\pm$75&8.00&Y&1, 3, 16\\
GG Tau Aa&4 32 30.35&+17 31 40.6&10100$\pm$7&J=9.07&K7$\pm$1&0.7&4060$\pm$250&4.50&Y&13, 17, 21\\
GG Tau Ab&4 32 30.35&+17 31 40.6&250.2$\pm$2.6&J=9.95&M0.5$\pm$0.5&3.2&3775$\pm$75&4.87&Y&13, 17, 21\\
GG Tau Ba&4 32 30.35&+17 31 40.6&10100$\pm$7&J=11.28&M6$\pm$0.5&0.6&2990$\pm$65&7.24&Y&13, 14, 17, 21\\
GG Tau Bb&4 32 30.35&+17 31 40.6&1476.5$\pm$6.5&J=12.96&M7.5$\pm$0.5&0.0&2795$\pm$85&9.10&Y&13, 14, 17, 21\\
UZ Tau Aa&4 32 43.04&+25 52 31.1&SB&H=8.54&M1$\pm$1&1.5&3705$\pm$145&4.77&Y&1, 18, 22\\
UZ Tau Ab&4 32 43.04&+25 52 31.1&SB&H=9.36&M4$\pm$1&1.5&3270$\pm$145&5.73&Y&1, 18, 22\\
UZ Tau Ba&4 32 43.04&+25 52 31.1&3539.5$\pm$2.1&H=8.46&M2$\pm$0.5&0.6&3560$\pm$75&4.91&Y&8, 9, 21\\
UZ Tau Bb&4 32 43.04&+25 52 31.1&367.8$\pm$1&H=9.18&M3$\pm$0.5&1.8&3415$\pm$75&5.48&Y&8, 9, 21\\
GH Tau A&4 33 06.22&+24 09 34.0&311.1$\pm$1.3&K=8.66&M2$\pm$0.5&0.0&3560$\pm$75&5.44&Y&8, 9, 21\\
GH Tau B&4 33 06.22&+24 09 34.0&311.1$\pm$1.3&K=8.45&M2$\pm$0.5&0.5&3560$\pm$75&5.17&Y&8, 9, 21\\
IS Tau A&4 33 36.79&+26 09 49.2&222.8$\pm$2.4&K=8.82&M0$\pm$0.5&3.3&3850$\pm$90&5.08&Y&8, 9\\
IS Tau B&4 33 36.79&+26 09 49.2&222.8$\pm$2.4&K=10.72&M3.5$\pm$0.5&3.6&3340$\pm$75&7.19&N&8, 9\\
HN Tau A&4 33 39.35&+17 51 52.4&3142$\pm$1&K=8.51&K5$\pm$1&0.5&4350$\pm$265&4.82&Y?&1, 13, 20\\
HN Tau B&4 33 39.35&+17 51 52.4&3142$\pm$1&K=10.81&M4.5$\pm$1&0.5&3200$\pm$180&7.72&Y?&4, 13, 20\\
IT Tau A&4 33 54.70&+26 13 27.5&2416$\pm$8&K=8.12&K3$\pm$1&4.1&4750$\pm$155&3.84&Y?&4, 13, 22\\
IT Tau B&4 33 54.70&+26 13 27.5&2416$\pm$8&K=9.54&M4$\pm$1&4.1&3270$\pm$145&5.98&Y?&4, 13, 22\\
Haro 6-28 A&4 35 56.84&+22 54 36.0&647$\pm$12&K=10.12&M2$\pm$0.5&2.3&3560$\pm$75&6.64&Y?&8, 9, 24\\
Haro 6-28 B&4 35 56.84&+22 54 36.0&647$\pm$12&K=10.48&M3.5$\pm$0.5&1.9&3340$\pm$75&7.14&Y?&8, 9, 24\\
2M04414565&4 41 45.65&+23 01 58.0&12370$\pm$70&J=10.74&M3$\pm$0.5&0.0&3415$\pm$75&6.72&...&3, 16\\
2M04414489&4 41 44.89&+23 01 51.3&12370$\pm$70&J=14.42&M8.25$\pm$0.25&0.0&2630$\pm$78&10.56&...&3, 16, 19\\
LkHa332-G2 A&4 42 07.33&+25 23 03.2&234.1$\pm$4.5&K=8.38&M0.5$\pm$0.5&2.0&3775$\pm$75&4.82&N&8, 9, 22\\
LkHa332-G2 B&4 42 07.33&+25 23 03.2&234.1$\pm$4.5&K=9.16&M2.5$\pm$0.5&3.3&3485$\pm$75&5.61&N&8, 9, 22\\
V955 Tau A&4 42 07.77&+25 23 11.8&330.9$\pm$1.2&K=8.18&K7$\pm$0.5&2.8&4060$\pm$125&4.26&Y&8, 9, 21\\
V955 Tau B&4 42 07.77&+25 23 11.8&330.9$\pm$1.2&K=9.72&M2.5$\pm$0.5&2.3&3485$\pm$75&6.28&Y&8, 9, 21\\
UY Aur A&4 51 47.38&+30 47 13.5&878$\pm$17&K=7.68&M0$\pm$0.5&0.6&3850$\pm$90&4.24&Y&8, 9\\
UY Aur B&4 51 47.38&+30 47 13.5&878$\pm$17&K=8.44&M2.5$\pm$0.5&2.7&3485$\pm$75&4.96&Y&8, 9\\
2M04554757&4 55 47.57&+30 28 07.7&6310$\pm$70&J=11.05&M4.75$\pm$0.25&0.0&3165$\pm$50&7.15&...&3, 7\\
2M04554801&4 55 48.01&+30 28 05.0&6310$\pm$70&J=13.19&M5.6$\pm$0.25&0.0&3045$\pm$35&9.31&...&3, 7\\
RW Aur A&5 07 49.54&+30 24 05.1&1417.5$\pm$3.4&K=7.25&K2$\pm$2&1.6&4900$\pm$330&3.11&Y&9, 10, 21\\
RW Aur B&5 07 49.54&+30 24 05.1&1417.5$\pm$3.4&K=8.82&K6$\pm$1&1.6&4350$\pm$265&5.01&Y&9, 10, 21\\
 \enddata
 \tablecomments{References: 
1) Kenyon \& Hartmann (1995), 
2) Leinert et al. (1993), 
3) Kraus et al. (in prep), 
4) Duch\^ene et al. (1999), 
5) White et al. (in prep), 
6) Boden et al. (2007), 
7) Luhman (2004), 
8) Hartigan \& Kenyon (2003), 
9) White \& Ghez (2001), 
10) White \& Hillenbrand (2004), 
11) Krist et al. (1998), 
12) Konopacky et al. (2007), 
13) Correia et al. (2006), 
14) White \& Basri (2003), 
15) Krist et al. (1995), 
16) Kraus \& Hillenbrand (2009), 
17) White et al. (1999), 
18) Prato et al. (2002), 
19) Luhman (2006), 
20) Furlan et al. (2006), 
21) McCabe et al. (2006), 
22) Hartmann et al. (2005), 
23) Luhman et al. (2006), 
24) Andrews \& Williams (2005).}
 \end{deluxetable*}

\clearpage

\begin{deluxetable*}{lrrrrrr}
\tablewidth{0pt}
\tabletypesize{\scriptsize}
\tablecaption{Derived Ages and Masses}
\tablehead{
\colhead{Name} & 
\colhead{$M$ ($M_{\sun}$)} & \colhead{log($\tau$) (yr)} &
\colhead{$M$ ($M_{\sun}$)} & \colhead{log($\tau$) (yr)} &
\colhead{$M$ ($M_{\sun}$)} & \colhead{log($\tau$) (yr)}
\\
\colhead{} &
\multicolumn{2}{c}{(Adopted)} & 
\multicolumn{2}{c}{(Lyon)} & 
\multicolumn{2}{c}{(DM97)}
}
\startdata
HBC 352&...&$>$7.7&...&$>$7.7&...&$>$7.7\\
HBC 353&...&$>$7.7&...&$>$7.7&...&$>$7.7\\
HBC 355&...&$>$7.7&...&$>$7.7&...&$>$7.7\\
HBC 354&...&$>$7.7&...&$>$7.7&...&$>$7.7\\
HBC 356&0.79$^{+0.08}_{-0.03}$&7.68$^{+0.65}_{-0.27}$&...&$>$7.7&0.79$^{+0.09}_{-0.05}$&7.56$^{+0.52}_{-0.27}$\\
HBC 357&0.79$^{+0.08}_{-0.03}$&7.68$^{+0.65}_{-0.27}$&...&$>$7.7&0.79$^{+0.09}_{-0.05}$&7.56$^{+0.52}_{-0.27}$\\
V773 Tau Aa&1.5$^{+0.4}_{-0.3}$&5.61$^{+0.22}_{-0.21}$&1.9$^{+0.3}_{-0.2}$&5.92$^{+0.28}_{-0.27}$&1.5$^{+0.4}_{-0.3}$&5.61$^{+0.22}_{-0.21}$\\
V773 Tau Ab&1.0$^{+0.5}_{-0.1}$&5.62$^{+0.40}_{-0.24}$&1.9$^{+0.3}_{-0.3}$&6.25$^{+0.48}_{-0.31}$&1.1$^{+0.4}_{-0.4}$&5.69$^{+0.33}_{-0.35}$\\
2M04141188&0.070$^{+0.008}_{-0.007}$&6.46$^{+0.19}_{-0.19}$&0.070$^{+0.008}_{-0.007}$&6.46$^{+0.19}_{-0.19}$&0.111$^{+0.013}_{-0.014}$&6.83$^{+0.11}_{-0.12}$\\
FO Tau A&0.4$^{+0.10}_{-0.17}$&5.86$^{+0.21}_{-0.28}$&0.40$^{+0.10}_{-0.17}$&5.86$^{+0.21}_{-0.28}$&0.19$^{+0.03}_{-0.03}$&5.20$^{+0.54}_{-0.27}$\\
FO Tau B&0.4$^{+0.10}_{-0.17}$&5.86$^{+0.21}_{-0.28}$&0.40$^{+0.10}_{-0.17}$&5.86$^{+0.21}_{-0.28}$&0.19$^{+0.03}_{-0.03}$&5.20$^{+0.54}_{-0.27}$\\
DD Tau A&0.43$^{+0.13}_{-0.23}$&5.63$^{+0.23}_{-0.32}$&0.43$^{+0.13}_{-0.23}$&5.63$^{+0.23}_{-0.32}$&0.18$^{+0.03}_{-0.02}$&4.85$^{+0.27}_{-0.25}$\\
DD Tau B&0.41$^{+0.11}_{-0.19}$&5.79$^{+0.22}_{-0.29}$&0.41$^{+0.11}_{-0.19}$&5.79$^{+0.22}_{-0.29}$&0.19$^{+0.03}_{-0.02}$&5.09$^{+0.35}_{-0.26}$\\
FQ Tau A&0.37$^{+0.07}_{-0.05}$&6.50$^{+0.23}_{-0.20}$&0.37$^{+0.07}_{-0.05}$&6.50$^{+0.23}_{-0.20}$&0.25$^{+0.05}_{-0.03}$&6.23$^{+0.20}_{-0.16}$\\
FQ Tau B&0.32$^{+0.05}_{-0.05}$&6.43$^{+0.20}_{-0.19}$&0.32$^{+0.05}_{-0.05}$&6.43$^{+0.20}_{-0.19}$&0.22$^{+0.03}_{-0.03}$&6.21$^{+0.17}_{-0.15}$\\
LkCa 7 A&0.67$^{+0.04}_{-0.04}$&6.28$^{+0.21}_{-0.21}$&0.96$^{+0.09}_{-0.12}$&6.61$^{+0.19}_{-0.22}$&0.42$^{+0.07}_{-0.06}$&5.91$^{+0.23}_{-0.18}$\\
LkCa 7 B&0.35$^{+0.06}_{-0.07}$&6.17$^{+0.17}_{-0.22}$&0.35$^{+0.06}_{-0.07}$&6.17$^{+0.17}_{-0.22}$&0.21$^{+0.04}_{-0.03}$&5.95$^{+0.15}_{-0.47}$\\
FS Tau A&0.66$^{+0.03}_{-0.03}$&5.67$^{+0.20}_{-0.20}$&1.06$^{+0.10}_{-0.12}$&6.06$^{+0.22}_{-0.25}$&0.32$^{+0.05}_{-0.04}$&5.16$^{+0.42}_{-0.39}$\\
FS Tau B&0.32$^{+0.05}_{-0.05}$&6.40$^{+0.20}_{-0.19}$&0.32$^{+0.05}_{-0.05}$&6.40$^{+0.20}_{-0.19}$&0.22$^{+0.03}_{-0.03}$&6.19$^{+0.17}_{-0.15}$\\
Haro 6-5B&...&$>$7.7&...&$>$7.7&...&$>$7.7\\
FV Tau A&0.83$^{+0.06}_{-0.06}$&5.85$^{+0.19}_{-0.18}$&1.54$^{+0.15}_{-0.15}$&6.38$^{+0.32}_{-0.22}$&0.56$^{+0.10}_{-0.08}$&5.61$^{+0.20}_{-0.17}$\\
FV Tau/c A&0.54$^{+0.05}_{-0.07}$&6.08$^{+0.19}_{-0.18}$&0.55$^{+0.07}_{-0.07}$&6.08$^{+0.18}_{-0.19}$&0.25$^{+0.04}_{-0.04}$&5.72$^{+0.16}_{-0.59}$\\
FV Tau/c B&0.30$^{+0.07}_{-0.05}$&6.67$^{+0.26}_{-0.20}$&0.30$^{+0.07}_{-0.05}$&6.67$^{+0.26}_{-0.20}$&0.23$^{+0.04}_{-0.03}$&6.48$^{+0.22}_{-0.19}$\\
DF Tau A&0.61$^{+0.19}_{-0.01}$&5.14$^{+0.22}_{-0.24}$&0.86$^{+0.06}_{-0.10}$&5.37$^{+0.17}_{-0.16}$&0.19$^{+0.02}_{-0.02}$&4.22$^{+0.25}_{-0.24}$\\
DF Tau B&0.65$^{+0.05}_{-0.08}$&5.74$^{+0.15}_{-0.27}$&0.65$^{+0.08}_{-0.08}$&5.74$^{+0.20}_{-0.20}$&0.22$^{+0.03}_{-0.02}$&4.83$^{+0.28}_{-0.24}$\\
2M04284263 A&0.16$^{+0.04}_{-0.04}$&6.50$^{+0.16}_{-0.19}$&0.16$^{+0.04}_{-0.04}$&6.50$^{+0.16}_{-0.19}$&0.17$^{+0.02}_{-0.02}$&6.52$^{+0.20}_{-0.17}$\\
2M04284263 B&0.098$^{+0.025}_{-0.019}$&6.62$^{+0.34}_{-0.23}$&0.098$^{+0.025}_{-0.019}$&6.62$^{+0.34}_{-0.23}$&0.148$^{+0.025}_{-0.024}$&6.91$^{+0.19}_{-0.18}$\\
UX Tau A&1.3$^{+0.3}_{-0.4}$&6.10$^{+0.26}_{-0.30}$&1.9$^{+0.3}_{-0.3}$&6.79$^{+0.19}_{-0.42}$&1.3$^{+0.3}_{-0.4}$&6.10$^{+0.26}_{-0.30}$\\
UX Tau C&0.16$^{+0.04}_{-0.04}$&6.51$^{+0.16}_{-0.19}$&0.16$^{+0.04}_{-0.04}$&6.51$^{+0.16}_{-0.19}$&0.17$^{+0.02}_{-0.02}$&6.54$^{+0.20}_{-0.17}$\\
FX Tau A&0.62$^{+0.05}_{-0.03}$&5.91$^{+0.20}_{-0.19}$&0.82$^{+0.19}_{-0.15}$&6.14$^{+0.29}_{-0.27}$&0.30$^{+0.07}_{-0.06}$&5.54$^{+0.23}_{-0.59}$\\
FX Tau B&0.28$^{+0.16}_{-0.32}$&5.90$^{+0.33}_{-1.04}$&0.28$^{+0.16}_{-0.32}$&5.90$^{+0.33}_{-1.04}$&0.18$^{+0.06}_{-0.04}$&5.41$^{+0.56}_{-0.38}$\\
DK Tau A&0.71$^{+0.09}_{-0.06}$&5.34$^{+0.32}_{-0.36}$&1.30$^{+0.16}_{-0.22}$&5.81$^{+0.32}_{-0.36}$&0.37$^{+0.13}_{-0.12}$&4.76$^{+0.65}_{-0.51}$\\
DK Tau B&0.61$^{+0.06}_{-0.04}$&6.17$^{+0.22}_{-0.20}$&0.78$^{+0.18}_{-0.17}$&6.38$^{+0.29}_{-0.27}$&0.35$^{+0.08}_{-0.07}$&5.81$^{+0.21}_{-0.23}$\\
V927 Tau A&0.42$^{+0.07}_{-0.06}$&6.21$^{+0.19}_{-0.18}$&0.42$^{+0.07}_{-0.06}$&6.21$^{+0.19}_{-0.18}$&0.24$^{+0.04}_{-0.04}$&5.93$^{+0.15}_{-0.31}$\\
V927 Tau B&0.33$^{+0.05}_{-0.05}$&6.32$^{+0.20}_{-0.18}$&0.33$^{+0.05}_{-0.05}$&6.32$^{+0.20}_{-0.18}$&0.22$^{+0.03}_{-0.03}$&6.12$^{+0.15}_{-0.15}$\\
HL Tau&0.83$^{+0.11}_{-0.10}$&5.93$^{+0.24}_{-0.25}$&1.53$^{+0.21}_{-0.27}$&6.51$^{+0.35}_{-0.35}$&0.57$^{+0.21}_{-0.15}$&5.68$^{+0.33}_{-0.22}$\\
XZ Tau A&0.59$^{+0.07}_{-0.04}$&5.86$^{+0.20}_{-0.23}$&0.67$^{+0.14}_{-0.12}$&5.97$^{+0.27}_{-0.28}$&0.25$^{+0.06}_{-0.05}$&5.16$^{+0.57}_{-0.36}$\\
XZ Tau B&0.46$^{+0.17}_{-0.30}$&5.36$^{+0.24}_{-0.36}$&0.46$^{+0.17}_{-0.30}$&5.36$^{+0.24}_{-0.36}$&0.17$^{+0.02}_{-0.01}$&4.44$^{+0.25}_{-0.25}$\\
HK Tau A&0.62$^{+0.03}_{-0.02}$&6.01$^{+0.19}_{-0.17}$&0.80$^{+0.10}_{-0.09}$&6.24$^{+0.2}_{-0.21}$&0.32$^{+0.05}_{-0.04}$&5.69$^{+0.15}_{-0.45}$\\
HK Tau B&...&$>$7.7&...&$>$7.7&...&$>$7.7\\
V710 Tau A&0.62$^{+0.09}_{-0.07}$&6.57$^{+0.24}_{-0.21}$&0.73$^{+0.15}_{-0.18}$&6.74$^{+0.26}_{-0.32}$&0.43$^{+0.12}_{-0.10}$&6.21$^{+0.31}_{-0.25}$\\
V710 Tau B&0.40$^{+0.14}_{-0.12}$&6.29$^{+0.32}_{-0.30}$&0.40$^{+0.14}_{-0.12}$&6.29$^{+0.32}_{-0.30}$&0.24$^{+0.10}_{-0.06}$&6.01$^{+0.22}_{-0.35}$\\
V710 Tau C&0.35$^{+0.07}_{-0.07}$&7.24$^{+0.21}_{-0.25}$&0.35$^{+0.07}_{-0.07}$&7.24$^{+0.21}_{-0.25}$&0.29$^{+0.05}_{-0.04}$&7.06$^{+0.24}_{-0.24}$\\
GG Tau Aa&0.73$^{+0.09}_{-0.08}$&5.88$^{+0.25}_{-0.30}$&1.21$^{+0.25}_{-0.22}$&6.34$^{+0.35}_{-0.33}$&0.43$^{+0.15}_{-0.12}$&5.62$^{+0.25}_{-0.57}$\\
GG Tau Ab&0.64$^{+0.03}_{-0.03}$&5.93$^{+0.19}_{-0.18}$&0.91$^{+0.10}_{-0.10}$&6.23$^{+0.20}_{-0.22}$&0.33$^{+0.05}_{-0.04}$&5.59$^{+0.18}_{-0.46}$\\
GG Tau Ba&0.104$^{+0.024}_{-0.020}$&5.80$^{+0.31}_{-0.51}$&0.104$^{+0.024}_{-0.020}$&5.80$^{+0.31}_{-0.51}$&0.134$^{+0.014}_{-0.012}$&6.11$^{+0.16}_{-0.44}$\\
GG Tau Bb&0.044$^{+0.011}_{-0.013}$&5.63$^{+0.61}_{-2.08}$&0.044$^{+0.011}_{-0.013}$&5.63$^{+0.61}_{-2.08}$&0.062$^{+0.021}_{-0.018}$&6.56$^{+0.14}_{-0.71}$\\
UZ Tau Aa&0.62$^{+0.05}_{-0.03}$&5.83$^{+0.19}_{-0.20}$&0.84$^{+0.18}_{-0.14}$&6.06$^{+0.29}_{-0.27}$&0.29$^{+0.07}_{-0.05}$&5.27$^{+0.43}_{-0.42}$\\
UZ Tau Ab&0.26$^{+0.20}_{-0.35}$&5.76$^{+0.38}_{-1.20}$&0.26$^{+0.20}_{-0.35}$&5.76$^{+0.38}_{-1.20}$&0.17$^{+0.06}_{-0.04}$&5.20$^{+0.67}_{-0.34}$\\
UZ Tau Ba&0.59$^{+0.05}_{-0.01}$&5.81$^{+0.19}_{-0.21}$&0.68$^{+0.07}_{-0.08}$&5.93$^{+0.19}_{-0.17}$&0.25$^{+0.04}_{-0.03}$&5.09$^{+0.53}_{-0.28}$\\
UZ Tau Bb&0.49$^{+0.07}_{-0.09}$&5.98$^{+0.18}_{-0.20}$&0.49$^{+0.07}_{-0.09}$&5.98$^{+0.18}_{-0.20}$&0.22$^{+0.04}_{-0.03}$&5.33$^{+0.53}_{-0.31}$\\
GH Tau A&0.58$^{+0.02}_{-0.05}$&6.15$^{+0.19}_{-0.20}$&0.61$^{+0.09}_{-0.07}$&6.20$^{+0.20}_{-0.18}$&0.29$^{+0.05}_{-0.04}$&5.77$^{+0.15}_{-0.47}$\\
GH Tau B&0.58$^{+0.02}_{-0.02}$&5.98$^{+0.20}_{-0.20}$&0.64$^{+0.08}_{-0.07}$&6.07$^{+0.18}_{-0.18}$&0.27$^{+0.04}_{-0.03}$&5.50$^{+0.29}_{-0.49}$\\
IS Tau A&0.67$^{+0.04}_{-0.03}$&6.11$^{+0.21}_{-0.20}$&0.99$^{+0.09}_{-0.12}$&6.46$^{+0.19}_{-0.21}$&0.39$^{+0.06}_{-0.05}$&5.77$^{+0.19}_{-0.18}$\\
IS Tau B&0.30$^{+0.06}_{-0.05}$&6.62$^{+0.25}_{-0.20}$&0.30$^{+0.06}_{-0.05}$&6.62$^{+0.25}_{-0.20}$&0.23$^{+0.04}_{-0.03}$&6.42$^{+0.21}_{-0.18}$\\
HN Tau A&0.85$^{+0.11}_{-0.10}$&6.27$^{+0.27}_{-0.25}$&1.35$^{+0.13}_{-0.16}$&6.85$^{+0.30}_{-0.34}$&0.65$^{+0.23}_{-0.18}$&6.05$^{+0.38}_{-0.32}$\\
HN Tau B&0.20$^{+0.13}_{-0.10}$&6.59$^{+0.45}_{-0.49}$&0.20$^{+0.13}_{-0.10}$&6.59$^{+0.45}_{-0.49}$&0.19$^{+0.07}_{-0.05}$&6.54$^{+0.33}_{-0.23}$\\
IT Tau A&1.0$^{+0.3}_{-0.1}$&5.96$^{+0.32}_{-0.21}$&1.8$^{+0.2}_{-0.2}$&6.76$^{+0.18}_{-0.39}$&1.0$^{+0.3}_{-0.3}$&5.96$^{+0.31}_{-0.32}$\\
IT Tau B&0.28$^{+0.15}_{-0.32}$&5.90$^{+0.33}_{-1.04}$&0.28$^{+0.15}_{-0.32}$&5.90$^{+0.33}_{-1.04}$&0.18$^{+0.06}_{-0.04}$&5.42$^{+0.55}_{-0.38}$\\
Haro 6-28 A&0.52$^{+0.07}_{-0.08}$&6.80$^{+0.21}_{-0.22}$&0.52$^{+0.07}_{-0.08}$&6.80$^{+0.21}_{-0.22}$&0.35$^{+0.07}_{-0.06}$&6.45$^{+0.24}_{-0.23}$\\
Haro 6-28 B&0.30$^{+0.06}_{-0.05}$&6.60$^{+0.24}_{-0.20}$&0.30$^{+0.06}_{-0.05}$&6.60$^{+0.24}_{-0.20}$&0.23$^{+0.04}_{-0.03}$&6.39$^{+0.21}_{-0.17}$\\
2M04414565&0.37$^{+0.08}_{-0.05}$&6.54$^{+0.24}_{-0.20}$&0.37$^{+0.08}_{-0.05}$&6.54$^{+0.24}_{-0.20}$&0.25$^{+0.05}_{-0.03}$&6.26$^{+0.21}_{-0.16}$\\
2M04414489&0.027$^{+0.006}_{-0.009}$&6.47$^{+0.44}_{-0.94}$&0.027$^{+0.006}_{-0.009}$&6.47$^{+0.44}_{-0.94}$&0.022$^{+0.007}_{-0.006}$&6$^{+0.82}_{-0.47}$\\
LkHa332-G2 A&0.64$^{+0.03}_{-0.02}$&5.90$^{+0.19}_{-0.18}$&0.92$^{+0.10}_{-0.10}$&6.20$^{+0.20}_{-0.21}$&0.33$^{+0.04}_{-0.04}$&5.56$^{+0.19}_{-0.50}$\\
LkHa332-G2 B&0.52$^{+0.05}_{-0.07}$&6.15$^{+0.19}_{-0.18}$&0.52$^{+0.07}_{-0.07}$&6.15$^{+0.19}_{-0.19}$&0.26$^{+0.05}_{-0.04}$&5.81$^{+0.14}_{-0.53}$\\
V955 Tau A&0.72$^{+0.05}_{-0.04}$&5.73$^{+0.22}_{-0.24}$&1.24$^{+0.11}_{-0.10}$&6.20$^{+0.22}_{-0.22}$&0.41$^{+0.07}_{-0.06}$&5.51$^{+0.15}_{-0.55}$\\
V955 Tau B&0.45$^{+0.08}_{-0.07}$&6.46$^{+0.23}_{-0.20}$&0.45$^{+0.08}_{-0.07}$&6.46$^{+0.23}_{-0.20}$&0.28$^{+0.06}_{-0.04}$&6.12$^{+0.20}_{-0.16}$\\
UY Aur A&0.66$^{+0.03}_{-0.02}$&5.56$^{+0.20}_{-0.20}$&1.07$^{+0.12}_{-0.12}$&5.95$^{+0.23}_{-0.25}$&0.31$^{+0.04}_{-0.04}$&4.94$^{+0.56}_{-0.31}$\\
UY Aur B&0.62$^{+0.05}_{-0.08}$&5.84$^{+0.15}_{-0.23}$&0.62$^{+0.08}_{-0.07}$&5.84$^{+0.19}_{-0.20}$&0.22$^{+0.03}_{-0.02}$&4.99$^{+0.43}_{-0.24}$\\
2M04554757&0.20$^{+0.03}_{-0.03}$&6.27$^{+0.17}_{-0.17}$&0.20$^{+0.03}_{-0.03}$&6.27$^{+0.17}_{-0.17}$&0.17$^{+0.02}_{-0.01}$&6.22$^{+0.14}_{-0.13}$\\
2M04554801&0.092$^{+0.008}_{-0.012}$&6.96$^{+0.76}_{-0.27}$&0.092$^{+0.008}_{-0.012}$&6.96$^{+0.76}_{-0.27}$&0.133$^{+0.015}_{-0.016}$&7.13$^{+0.16}_{-0.15}$\\
RW Aur A&1.4$^{+0.6}_{-0.7}$&5.85$^{+0.44}_{-0.53}$&2.1$^{+0.3}_{-0.5}$&6.32$^{+0.54}_{-0.42}$&1.4$^{+0.6}_{-0.7}$&5.85$^{+0.44}_{-0.53}$\\
RW Aur B&0.86$^{+0.11}_{-0.10}$&6.40$^{+0.26}_{-0.26}$&1.26$^{+0.14}_{-0.14}$&6.96$^{+0.29}_{-0.33}$&0.69$^{+0.24}_{-0.19}$&6.19$^{+0.39}_{-0.35}$\\
\enddata
\end{deluxetable*}

\clearpage

\begin{deluxetable*}{lllllrlrrrl}
\tablewidth{0pt}
\tabletypesize{\scriptsize}
\tablecaption{Single Star Sample: Observed Properties}
\tablehead{
\colhead{Name} & \colhead{RA} & \colhead{Dec} &
\colhead{$J$} & \colhead{SpT} & \colhead{$A_V$} &
\colhead{$T_{eff}$} & \colhead{$M_{bol}$} & \colhead{$M$} & \colhead{$\log(\tau)$} & \colhead{Refs}
\\
\colhead{} & \multicolumn{2}{c}{(J2000)} &
\colhead{(mag)} & \colhead{} & \colhead{(mag)} & \colhead{(K)} & \colhead{(mag)} & \colhead{($M_{\sun}$)} & \colhead{(yr)}
}
\startdata
2MASSJ04080782+2807280&4 08 07.82&+28 07 28.0&12.44&M3.75$\pm$0.25&1.0&3305$\pm$35&8.18&0.25$^{+0.03}_{-0.03}$&7.04$^{+0.16}_{-0.17}$&1, 2\\
LkCa1&4 13 14.14&+28 19 10.8&9.64&M4$\pm$0.5&0.0&3270$\pm$70&5.66&0.26$^{+0.13}_{-0.20}$&5.70$^{+0.28}_{-0.32}$&3, 4\\
Anon1&4 13 27.23&+28 16 24.8&8.83&M0$\pm$1&3.6&3850$\pm$178&3.59&0.65$^{+0.04}_{-0.03}$&5.13$^{+0.20}_{-0.21}$&3, 5, 6\\
2MASSJ04141188+2811535&4 14 11.88&+28 11 53.5&13.16&M6.25$\pm$0.25&0.7&2960$\pm$28&9.08&0.069$^{+0.008}_{-0.006}$&6.50$^{+0.19}_{-0.18}$&1, 7\\
FMTau&4 14 13.58&+28 12 49.2&10.33&M0$\pm$1&1.9&3850$\pm$178&5.56&0.68$^{+0.09}_{-0.08}$&6.42$^{+0.23}_{-0.24}$&3, 4, 6\\
FNTau&4 14 14.59&+28 27 58.1&9.47&M5$\pm$0.5&1.4&3125$\pm$70&5.23&-0.20$^{+0.20}_{-1.10}$&4.20$^{+0.97}_{-22.11}$&3, 8\\
CWTau&4 14 17.00&+28 10 57.8&9.56&K3$\pm$1&1.9&4730$\pm$155&4.60&1.00$^{+0.22}_{-0.07}$&6.39$^{+0.29}_{-0.21}$&3, 6, 8\\
CIDA-1&4 14 17.61&+28 06 09.7&11.73&M5.5$\pm$0.5&3.0&3060$\pm$68&7.02&0.02$^{+0.15}_{-0.23}$&4.40$^{+1.80}_{-3.98}$&9, 10, 11\\
MHO-1&4 14 26.40&+28 05 59.7&11.52&M2.5$\pm$0.5&5.7&3490$\pm$73&5.90&0.49$^{+0.07}_{-0.07}$&6.30$^{+0.20}_{-0.18}$&6, 10, 12\\
FPTau&4 14 47.31&+26 46 26.4&9.89&M4$\pm$0.5&0.2&3270$\pm$70&5.85&0.27$^{+0.11}_{-0.18}$&5.82$^{+0.26}_{-0.31}$&3, 4\\
CXTau&4 14 47.86&+26 48 11.0&9.87&M2.5$\pm$0.5&0.8&3490$\pm$73&5.59&0.53$^{+0.05}_{-0.07}$&6.15$^{+0.19}_{-0.18}$&3, 4\\
KPNO-Tau-1&4 15 14.71&+28 00 09.6&15.10&M8.5$\pm$0.25&0.4&2555$\pm$78&11.13&0.023$^{+0.010}_{-0.006}$&6.75$^{+0.40}_{-1.36}$&13, 14, 15\\
2MASSJ04152409+2910434&4 15 24.09&+29 10 43.4&13.68&M7$\pm$0.25&2.0&2880$\pm$33&9.27&0.053$^{+0.006}_{-0.005}$&6.31$^{+0.26}_{-0.26}$&1, 2\\
2MASSJ04161210+2756386&4 16 12.10&+27 56 38.6&12.27&M4.75$\pm$0.25&2.0&3160$\pm$38&7.81&0.181$^{+0.022}_{-0.023}$&6.55$^{+0.11}_{-0.14}$&7, 16\\
2MASSJ04161885+2752155&4 16 18.85&+27 52 15.5&12.55&M6.25$\pm$0.25&1.0&2960$\pm$28&8.39&0.080$^{+0.008}_{-0.008}$&6.18$^{+0.16}_{-0.16}$&1, 2\\
2MASSJ04163911+2858491&4 16 39.11&+28 58 49.1&12.72&M5.5$\pm$0.25&2.8&3060$\pm$38&8.07&0.13$^{+0.02}_{-0.03}$&6.44$^{+0.12}_{-0.18}$&1, 2\\
CYTau&4 17 33.73&+28 20 46.9&9.83&M1$\pm$0.5&0.1&3705$\pm$73&5.64&0.61$^{+0.04}_{-0.02}$&6.35$^{+0.20}_{-0.20}$&3, 8\\
KPNO-Tau-10&4 17 49.55&+28 13 31.9&11.89&M5$\pm$0.25&0.0&3125$\pm$33&8.02&0.15$^{+0.02}_{-0.02}$&6.55$^{+0.11}_{-0.13}$&1, 14, 17\\
V410-Xray1&4 17 49.65&+28 29 36.3&11.02&M3.75$\pm$0.25&0.9&3305$\pm$35&6.78&0.29$^{+0.03}_{-0.02}$&6.36$^{+0.16}_{-0.14}$&6, 10, 18\\
V410-Anon13&4 18 17.11&+28 28 41.9&12.96&M6$\pm$0.5&3.8&2990$\pm$63&8.02&0.09$^{+0.03}_{-0.02}$&6.12$^{+0.31}_{-0.25}$&6, 12, 15\\
KPNO-Tau-11&4 18 30.31&+27 43 20.8&11.89&M5.5$\pm$0.25&0.0&3060$\pm$38&8.01&0.13$^{+0.02}_{-0.03}$&6.42$^{+0.12}_{-0.17}$&1, 17\\
KPNO-Tau-2&4 18 51.16&+28 14 33.2&13.92&M7.5$\pm$0.25&0.4&2795$\pm$45&9.96&0.040$^{+0.007}_{-0.005}$&6.71$^{+0.13}_{-0.67}$&13, 14, 15\\
HBC376&4 18 51.70&+17 23 16.6&10.03&K7$\pm$1&0.0&4060$\pm$250&5.66&0.79$^{+0.10}_{-0.13}$&6.63$^{+0.27}_{-0.26}$&3, 4\\
I04158+2805&4 18 58.14&+28 12 23.5&13.78&M6$\pm$1&8.6&2990$\pm$123&7.50&0.10$^{+0.07}_{-0.04}$&5.91$^{+0.45}_{-0.88}$&10, 19\\
KPNO-Tau-12&4 19 01.27&+28 02 48.7&16.31&M9$\pm$0.25&0.5&2400$\pm$75&12.31&0.032$^{+0.007}_{-0.010}$&7.46$^{+0.23}_{-0.28}$&15, 17\\
V410-Xray5a&4 19 01.98&+28 22 33.2&11.99&M5.5$\pm$0.5&2.6&3060$\pm$68&7.41&0.14$^{+0.04}_{-0.12}$&6.20$^{+0.17}_{-1.88}$&6, 10, 12\\
BPTau&4 19 15.84&+29 06 26.9&9.10&K7$\pm$1&0.5&4060$\pm$250&4.59&0.73$^{+0.09}_{-0.08}$&5.94$^{+0.25}_{-0.29}$&3, 5\\
2MASSJ04202555+2700355&4 20 25.55&+27 00 35.5&12.86&M5.25$\pm$0.25&2.0&3095$\pm$33&8.42&0.123$^{+0.018}_{-0.014}$&6.62$^{+0.11}_{-0.11}$&1, 7\\
J2-157&4 20 52.73&+17 46 41.5&11.62&M5.5$\pm$0.5&0.0&3060$\pm$68&7.74&0.14$^{+0.03}_{-0.04}$&6.32$^{+0.17}_{-0.35}$&3, 10\\
CFHT-Tau-19&4 21 07.95&+27 02 20.4&13.85&M5.25$\pm$0.25&7.3&3095$\pm$33&7.95&0.145$^{+0.020}_{-0.018}$&6.46$^{+0.12}_{-0.13}$&14, 16\\
2MASSJ04213460+2701388&4 21 34.60&+27 01 38.8&11.90&M5.5$\pm$0.25&1.8&3060$\pm$38&7.53&0.14$^{+0.02}_{-0.05}$&6.24$^{+0.13}_{-0.73}$&7, 16\\
CFHT-Tau-10&4 21 46.31&+26 59 29.6&13.82&M5.75$\pm$0.25&2.0&3020$\pm$35&9.38&0.084$^{+0.012}_{-0.012}$&6.88$^{+0.45}_{-0.22}$&1, 2\\
2MASSJ04215450+2652315&4 21 54.50&+26 52 31.5&15.54&M8.5$\pm$0.25&1.0&2555$\pm$78&11.40&0.027$^{+0.012}_{-0.009}$&7.09$^{+0.23}_{-0.85}$&1, 2\\
DETau&4 21 55.64&+27 55 06.1&9.18&M2$\pm$0.5&0.6&3560$\pm$70&4.93&0.59$^{+0.04}_{-0.01}$&5.82$^{+0.19}_{-0.20}$&3, 8\\
RYTau&4 21 57.40&+28 26 35.5&7.16&K1$\pm$1&1.8&5080$\pm$175&2.09&2.1$^{+0.6}_{-0.6}$&5.61$^{+0.28}_{-0.35}$&3, 8\\
HD283572&4 21 58.84&+28 18 06.6&7.42&G5$\pm$2&0.4&5770$\pm$98&2.54&2.0$^{+0.4}_{-0.2}$&6.62$^{+0.10}_{-0.25}$&3, 5\\
CFHT-Tau-14&4 22 16.44&+25 49 11.8&13.06&M7.75$\pm$0.25&0.5&2750$\pm$43&9.06&0.036$^{+0.005}_{-0.006}$&4.19$^{+1.12}_{-2.14}$&1, 2, 14\\
CFHT-Tau-21&4 22 16.76&+26 54 57.1&11.58&M1.5$\pm$0.25&3.0&3630$\pm$38&6.63&0.58$^{+0.02}_{-0.03}$&6.91$^{+0.18}_{-0.18}$&2, 16\\
2MJ04230607+2801194&4 23 06.07&+28 01 19.4&12.24&M6.25$\pm$0.25&0.0&2960$\pm$28&8.36&0.080$^{+0.008}_{-0.008}$&6.17$^{+0.16}_{-0.16}$&1, 20\\
CFHT-Tau-9&4 24 26.46&+26 49 50.4&12.88&M5.75$\pm$0.25&0.5&3020$\pm$35&8.86&0.087$^{+0.012}_{-0.010}$&6.57$^{+0.21}_{-0.17}$&1, 2\\
IPTau&4 24 57.08&+27 11 56.5&9.78&M0$\pm$1&0.2&3850$\pm$178&5.47&0.68$^{+0.08}_{-0.08}$&6.37$^{+0.23}_{-0.24}$&3, 5\\
KPNO-Tau-3&4 26 29.39&+26 24 13.8&13.32&M6$\pm$0.25&1.6&2990$\pm$30&8.98&0.077$^{+0.009}_{-0.008}$&6.55$^{+0.17}_{-0.17}$&13, 15\\
KPNO-Tau-13&4 26 57.33&+26 06 28.4&11.28&M5$\pm$0.25&2.5&3125$\pm$33&6.71&0.12$^{+0.08}_{-0.07}$&5.84$^{+0.31}_{-0.51}$&10, 17\\
HBC388&4 27 10.56&+17 50 42.6&8.79&K1$\pm$1&0.1&5080$\pm$175&4.20&1.42$^{+0.12}_{-0.13}$&6.63$^{+0.24}_{-0.26}$&3, 5\\
KPNO-Tau-4&4 27 28.00&+26 12 05.3&15.00&M9.5$\pm$0.25&2.5&2245$\pm$80&10.46&-0.015$^{+0.011}_{-0.006}$&3.45$^{+1.41}_{-0.58}$&13, 14, 15\\
2MASSJ04290068+2755033&4 29 00.68&+27 55 03.3&14.02&M8.25$\pm$0.25&0.0&2630$\pm$578&10.16&0.02$^{+0.13}_{-0.17}$&4.46$^{+3.38}_{-16.38}$&1, 2\\
KPNO-Tau-5&4 29 45.68&+26 30 46.8&12.64&M7.5$\pm$0.25&0.0&2795$\pm$45&8.78&0.043$^{+0.008}_{-0.006}$&4.77$^{+1.00}_{-1.44}$&13, 15\\
IQTau&4 29 51.56&+26 06 44.9&9.42&M0.5$\pm$0.5&1.3&3775$\pm$73&4.87&0.64$^{+0.03}_{-0.02}$&5.93$^{+0.19}_{-0.18}$&3, 5\\
CFHT-Tau-20&4 29 59.51&+24 33 07.9&11.68&M5$\pm$0.25&2.2&3125$\pm$33&7.19&0.18$^{+0.02}_{-0.04}$&6.21$^{+0.15}_{-0.23}$&2, 16\\
KPNO-Tau-6&4 30 07.24&+26 08 20.8&15.00&M8.5$\pm$0.25&0.9&2555$\pm$78&10.90&0.021$^{+0.007}_{-0.007}$&5.98$^{+0.99}_{-1.92}$&13, 14, 15\\
CFHT-Tau-16&4 30 23.65&+23 59 13.0&14.96&M8.25$\pm$0.25&0.0&2630$\pm$578&11.10&0.03$^{+0.12}_{-0.11}$&7.00$^{+1.37}_{-11.40}$&1, 2\\
KPNO-Tau-7&4 30 57.19&+25 56 39.5&14.52&M8.25$\pm$0.25&0.0&2630$\pm$578&10.66&0.03$^{+0.12}_{-0.14}$&6.55$^{+1.53}_{-14.43}$&13, 15\\
JH56&4 31 14.44&+27 10 18.0&9.70&M0.5$\pm$0.5&1.1&3775$\pm$73&5.21&0.64$^{+0.03}_{-0.03}$&6.13$^{+0.20}_{-0.19}$&3, 10\\
MHO-9&4 31 15.78&+18 20 07.2&11.21&M5$\pm$0.5&2.2&3125$\pm$70&6.73&0.10$^{+0.10}_{-0.40}$&5.86$^{+0.34}_{-6.59}$&6, 9, 10\\
2MASSJ04311907+2335047&4 31 19.07&+23 35 04.7&13.51&M7.75$\pm$0.25&0.5&2750$\pm$43&9.51&0.036$^{+0.005}_{-0.005}$&5.28$^{+0.90}_{-1.61}$&1, 2\\
MHO-4&4 31 24.06&+18 00 21.5&11.66&M7$\pm$0.5&1.0&2880$\pm$70&7.53&0.07$^{+0.03}_{-0.02}$&5.19$^{+0.65}_{-1.44}$&6, 9, 15\\
CFHT-Tau-13&4 31 26.69&+27 03 18.8&14.83&M7.5$\pm$0.25&0.5&2795$\pm$45&10.84&0.049$^{+0.011}_{-0.010}$&7.14$^{+0.27}_{-0.21}$&1, 2\\
LkHa358&4 31 36.13&+18 13 43.3&12.80&M5.5$\pm$0.5&13.5&3060$\pm$68&5.18&-0.9$^{+1.0}_{-0.5}$&-9.28$^{+14.39}_{-8.92}$&3, 6, 10\\
HLTau&4 31 38.44&+18 13 57.7&10.62&K5$\pm$1&7.4&4350$\pm$265&4.25&0.83$^{+0.11}_{-0.10}$&5.93$^{+0.24}_{-0.25}$&8, 19\\
J1-665&4 31 58.44&+25 43 29.9&10.59&M5$\pm$0.5&1.0&3125$\pm$70&6.45&0.10$^{+0.10}_{-0.60}$&5.55$^{+0.50}_{-9.46}$&3, 10\\
2MASSJ04320329+2528078&4 32 03.29&+25 28 07.8&11.72&M6.25$\pm$0.25&0.0&2960$\pm$28&7.84&0.089$^{+0.008}_{-0.008}$&5.95$^{+0.16}_{-0.15}$&1, 2\\
L1551-51&4 32 09.27&+17 57 22.8&9.70&K7$\pm$1&0.0&4060$\pm$250&5.33&0.76$^{+0.11}_{-0.11}$&6.42$^{+0.25}_{-0.26}$&3, 4, 6\\
Haro6-13&4 32 15.41&+24 28 59.7&11.24&M0$\pm$0.5&11.9&3850$\pm$90&3.72&0.653$^{+0.018}_{-0.017}$&5.22$^{+0.20}_{-0.20}$&8, 19\\
MHO-5&4 32 16.07&+18 12 46.4&11.07&M7$\pm$0.5&0.1&2880$\pm$70&7.18&0.07$^{+0.03}_{-0.02}$&4.97$^{+0.75}_{-1.63}$&6, 9, 15\\
CFHT-Tau-7&4 32 17.86&+24 22 15.0&11.54&M5.75$\pm$0.25&0.0&3020$\pm$35&7.65&0.12$^{+0.02}_{-0.03}$&6.21$^{+0.13}_{-0.98}$&2, 10\\
MHO-6&4 32 22.11&+18 27 42.6&11.71&M4.75$\pm$0.25&0.9&3160$\pm$38&7.57&0.19$^{+0.02}_{-0.02}$&6.46$^{+0.14}_{-0.16}$&6, 10, 13\\
2MASSJ04322329+2403013&4 32 23.29&+24 03 01.3&12.34&M7.75$\pm$0.25&0.0&2750$\pm$43&8.48&0.037$^{+0.006}_{-0.006}$&2.76$^{+1.44}_{-2.87}$&1, 2\\
MHO-7&4 32 26.28&+18 27 52.1&11.11&M5.25$\pm$0.25&0.4&3095$\pm$33&7.12&0.15$^{+0.03}_{-0.08}$&6.12$^{+0.16}_{-0.99}$&6, 10, 13\\
FYTau&4 32 30.58&+24 19 57.3&9.98&K7$\pm$1&3.5&4060$\pm$250&4.65&0.74$^{+0.09}_{-0.09}$&5.98$^{+0.25}_{-0.29}$&3, 4\\
FZTau&4 32 31.76&+24 20 03.0&9.90&M0$\pm$1&3.6&3850$\pm$178&4.67&0.66$^{+0.06}_{-0.05}$&5.84$^{+0.24}_{-0.20}$&8, 13, 21\\
L1551-55&4 32 43.73&+18 02 56.3&10.16&K7$\pm$1&0.7&4060$\pm$250&5.60&0.78$^{+0.11}_{-0.12}$&6.59$^{+0.27}_{-0.26}$&3, 22\\
KPNO-Tau-14&4 33 07.81&+26 16 06.6&11.91&M6$\pm$0.25&3.1&2990$\pm$30&7.17&0.11$^{+0.01}_{-0.07}$&5.77$^{+0.16}_{-2.31}$&15, 17\\
V830Tau&4 33 10.03&+24 33 43.4&9.32&K7$\pm$1&0.3&4060$\pm$250&4.87&0.74$^{+0.10}_{-0.09}$&6.12$^{+0.26}_{-0.27}$&3, 5\\
I04303+2240&4 33 19.07&+22 46 34.2&11.103&M0.5$\pm$1&11.7&3775$\pm$163&3.67&0.64$^{+0.03}_{-0.03}$&5.20$^{+0.18}_{-0.19}$&10, 19\\
GITau&4 33 34.06&+24 21 17.0&9.34&K6$\pm$1&0.9&4350$\pm$265&4.78&0.85$^{+0.11}_{-0.10}$&6.25$^{+0.27}_{-0.25}$&3, 8\\
DLTau&4 33 39.06&+25 20 38.2&9.63&K7$\pm$1&1.7&4060$\pm$250&4.79&0.74$^{+0.09}_{-0.09}$&6.07$^{+0.25}_{-0.28}$&3, 4, 23\\
2MASSJ04334291+2526470&4 33 42.91&+25 26 47.0&14.64&M8.75$\pm$0.25&0.0&2475$\pm$78&10.78&0.014$^{+0.006}_{-0.007}$&5.74$^{+0.33}_{-1.48}$&1, 2\\
DMTau&4 33 48.72&+18 10 10.0&10.44&M1$\pm$0.5&0.0&3705$\pm$73&6.28&0.63$^{+0.05}_{-0.04}$&6.76$^{+0.20}_{-0.20}$&3, 4\\
CITau&4 33 52.00&+22 50 30.2&9.48&K7$\pm$1&1.8&4060$\pm$250&4.62&0.74$^{+0.09}_{-0.08}$&5.96$^{+0.25}_{-0.29}$&3, 8\\
JH108&4 34 10.99&+22 51 44.5&10.60&M1$\pm$0.5&1.5&3705$\pm$73&6.03&0.62$^{+0.05}_{-0.04}$&6.60$^{+0.20}_{-0.20}$&3, 24\\
CFHT-Tau-1&4 34 15.27&+22 50 31.0&13.74&M7$\pm$0.25&3.1&2880$\pm$33&9.02&0.055$^{+0.006}_{-0.005}$&6.14$^{+0.25}_{-0.27}$&15, 25\\
AATau&4 34 55.42&+24 28 53.2&9.44&K7$\pm$1&0.5&4060$\pm$250&4.93&0.74$^{+0.10}_{-0.09}$&6.16$^{+0.26}_{-0.27}$&3, 4\\
HOTau&4 35 20.20&+22 32 14.6&11.20&M0.5$\pm$0.5&1.1&3775$\pm$73&6.69&0.69$^{+0.04}_{-0.06}$&7.10$^{+0.23}_{-0.22}$&3, 24\\
DNTau&4 35 27.37&+24 14 58.9&9.14&M0$\pm$0.5&1.9&3850$\pm$90&4.37&0.66$^{+0.03}_{-0.03}$&5.65$^{+0.20}_{-0.20}$&5, 19\\
KPNO-Tau-8&4 35 41.84&+22 34 11.6&12.95&M5.75$\pm$0.25&0.5&3020$\pm$35&8.92&0.086$^{+0.011}_{-0.010}$&6.60$^{+0.24}_{-0.17}$&13, 15\\
KPNO-Tau-9&4 35 51.43&+22 49 11.9&15.48&M8.5$\pm$0.25&0.0&2555$\pm$78&11.62&0.033$^{+0.011}_{-0.013}$&7.24$^{+0.25}_{-0.43}$&13, 15\\
HPTau-G2&4 35 54.15&+22 54 13.5&8.10&G0$\pm$2&2.1&6030$\pm$170&2.67&1.83$^{+0.07}_{-0.07}$&6.75$^{+0.04}_{-0.05}$&3, 4\\
CFHT-Tau-2&4 36 10.39&+22 59 56.0&13.76&M7.5$\pm$0.25&2.0&2795$\pm$45&9.34&0.041$^{+0.007}_{-0.005}$&5.68$^{+0.63}_{-1.03}$&13, 15\\
LkCa14&4 36 19.09&+25 42 59.0&9.34&M0$\pm$1&0.0&3850$\pm$178&5.10&0.67$^{+0.07}_{-0.06}$&6.12$^{+0.24}_{-0.22}$&3, 4\\
CFHT-Tau-3&4 36 38.94&+22 58 11.9&13.73&M7.75$\pm$0.25&1.0&2750$\pm$43&9.59&0.036$^{+0.004}_{-0.005}$&5.46$^{+0.86}_{-1.53}$&13, 15\\
2MASSJ04380084+2558572&4 38 00.84&+25 58 57.2&11.54&M7.25$\pm$0.25&0.6&2840$\pm$43&7.51&0.059$^{+0.010}_{-0.010}$&4.72$^{+0.52}_{-1.82}$&7, 16\\
GMTau&4 38 21.34&+26 09 13.7&12.80&M6.5$\pm$0.5&2.0&2935$\pm$55&8.38&0.074$^{+0.015}_{-0.013}$&6.07$^{+0.25}_{-0.38}$&9, 11, 15\\
DOTau&4 38 28.58&+26 10 49.4&9.47&M0$\pm$1&2.6&3850$\pm$178&4.50&0.66$^{+0.06}_{-0.05}$&5.73$^{+0.23}_{-0.20}$&3, 8\\
SCHJ0439016+2336030&4 39 01.60&+23 36 03.0&11.34&M6$\pm$0.25&0.0&2990$\pm$30&7.45&0.10$^{+0.01}_{-0.03}$&5.89$^{+0.25}_{-1.02}$&1, 11, 26\\
CIDA-13&4 39 15.86&+30 32 07.4&12.68&M3.5$\pm$0.5&0.4&3340$\pm$73&8.59&0.25$^{+0.05}_{-0.05}$&7.31$^{+0.22}_{-0.23}$&10, 27\\
LkCa15&4 39 17.80&+22 21 03.5&9.42&K5$\pm$1&0.6&4350$\pm$265&4.93&0.85$^{+0.11}_{-0.10}$&6.35$^{+0.26}_{-0.26}$&3, 4\\
CFHT-Tau-4&4 39 47.48&+26 01 40.8&12.17&M7$\pm$0.25&3.0&2880$\pm$33&7.48&0.068$^{+0.012}_{-0.007}$&5.16$^{+0.39}_{-0.36}$&13, 15, 25\\
I04370+2559&4 40 08.00&+26 05 25.4&12.41&M4.75$\pm$0.25&10.0&3160$\pm$38&5.75&-0.05$^{+0.13}_{-0.09}$&5.24$^{+0.35}_{-0.60}$&2, 10\\
I04385+2550&4 41 38.82&+25 56 26.8&11.85&M0.5$\pm$0.5&10.2&3775$\pm$73&4.83&0.64$^{+0.03}_{-0.02}$&5.91$^{+0.19}_{-0.18}$&10, 19\\
CIDA-7&4 42 21.02&+25 20 34.4&11.40&M4.75$\pm$0.25&1.0&3160$\pm$38&7.22&0.199$^{+0.021}_{-0.023}$&6.29$^{+0.16}_{-0.15}$&2, 10\\
DPTau&4 42 37.70&+25 15 37.5&11.00&M0$\pm$1&6.3&3850$\pm$178&5.02&0.67$^{+0.07}_{-0.06}$&6.07$^{+0.24}_{-0.22}$&4, 19\\
GOTau&4 43 03.09&+25 20 18.8&10.71&M0$\pm$1&1.2&3850$\pm$178&6.15&0.72$^{+0.09}_{-0.11}$&6.81$^{+0.26}_{-0.25}$&3, 4\\
2MASSJ04442713+2512164&4 44 27.13&+25 12 16.4&12.19&M7.25$\pm$0.25&0.0&2840$\pm$43&8.33&0.054$^{+0.008}_{-0.008}$&5.37$^{+0.41}_{-1.22}$&7, 16\\
DQTau&4 46 53.05&+17 00 00.2&9.51&M0$\pm$1&1.0&3850$\pm$178&5.00&0.67$^{+0.07}_{-0.06}$&6.06$^{+0.24}_{-0.22}$&3, 4\\
DRTau&4 47 06.21&+16 58 42.8&8.84&K7$\pm$1&3.2&4060$\pm$250&3.59&0.70$^{+0.09}_{-0.06}$&5.29$^{+0.32}_{-0.36}$&3, 5, 23\\
DSTau&4 47 48.59&+29 25 11.2&9.47&K5$\pm$1&0.3&4350$\pm$265&5.06&0.86$^{+0.11}_{-0.11}$&6.44$^{+0.25}_{-0.26}$&3, 4\\
GMAur&4 55 10.98&+30 21 59.5&9.34&K3$\pm$1&0.1&4730$\pm$155&4.88&1.02$^{+0.16}_{-0.08}$&6.58$^{+0.26}_{-0.23}$&3, 4\\
2MASSJ04552333+3027366&4 55 23.33&+30 27 36.6&13.07&M6.25$\pm$0.25&0.0&2960$\pm$28&9.19&0.068$^{+0.007}_{-0.006}$&6.56$^{+0.19}_{-0.19}$&1, 7\\
LkCa19&4 55 36.96&+30 17 55.3&8.87&K0$\pm$2&0.0&5250$\pm$335&4.27&1.35$^{+0.19}_{-0.16}$&6.84$^{+0.27}_{-0.38}$&3, 4\\
2MASSJ04554046+3039057&4 55 40.46&+30 39 05.7&12.71&M5.25$\pm$0.25&0.3&3095$\pm$33&8.76&0.110$^{+0.015}_{-0.011}$&6.72$^{+0.22}_{-0.10}$&1, 7\\
2MASSJ04554535+3019389&4 55 45.35&+30 19 38.9&11.44&M4.75$\pm$0.25&0.0&3160$\pm$38&7.54&0.192$^{+0.020}_{-0.021}$&6.44$^{+0.14}_{-0.16}$&7, 16\\
2MASSJ04554970+3019400&4 55 49.70&+30 19 40.0&12.81&M6$\pm$0.25&0.0&2990$\pm$30&8.92&0.078$^{+0.009}_{-0.008}$&6.52$^{+0.17}_{-0.17}$&1, 7\\
2MASSJ04555289+3006523&4 55 52.89&+30 06 52.3&11.64&M5.25$\pm$0.25&0.0&3095$\pm$33&7.76&0.152$^{+0.019}_{-0.018}$&6.40$^{+0.12}_{-0.13}$&1, 7\\
2MASSJ04555637+3049375&4 55 56.37&+30 49 37.5&12.00&M5$\pm$0.25&0.4&3125$\pm$33&8.03&0.15$^{+0.02}_{-0.02}$&6.56$^{+0.11}_{-0.13}$&1, 7\\
SUAur&4 55 59.38&+30 34 01.6&7.20&G2$\pm$2&0.9&5860$\pm$115&2.11&2.3$^{+0.3}_{-0.3}$&6.39$^{+0.19}_{-0.21}$&3, 8\\
2MASSJ04574903+3015195&4 57 49.03&+30 15 19.5&15.77&M9.25$\pm$0.25&0.0&2325$\pm$78&11.91&0.016$^{+0.011}_{-0.005}$&6.54$^{+0.76}_{-0.55}$&1, 7\\
V836Tau&5 03 06.60&+25 23 19.7&9.92&K7$\pm$1&1.7&4060$\pm$250&5.08&0.75$^{+0.10}_{-0.10}$&6.26$^{+0.26}_{-0.26}$&4, 19\\
CIDA-8&5 04 41.40&+25 09 54.4&10.92&M3.5$\pm$0.5&3.0&3340$\pm$73&6.09&0.36$^{+0.06}_{-0.08}$&6.13$^{+0.17}_{-0.23}$&3, 10\\
CIDA-10&5 06 16.75&+24 46 10.2&10.80&M4$\pm$0.5&0.5&3270$\pm$70&6.68&0.27$^{+0.04}_{-0.04}$&6.25$^{+0.19}_{-0.20}$&3, 10\\
RX05072+2437&5 07 12.07&+24 37 16.4&10.14&K6$\pm$0.5&0.9&4350$\pm$133&5.56&0.90$^{+0.04}_{-0.06}$&6.77$^{+0.25}_{-0.22}$&10, 13, 27\\
CIDA-12&5 07 54.97&+25 00 15.6&11.42&M4$\pm$0.5&0.8&3270$\pm$70&7.22&0.26$^{+0.04}_{-0.04}$&6.51$^{+0.19}_{-0.20}$&3, 10\\
\enddata
\tablenotetext{a}{Some systems which sit extremely high or low in the HR diagram exceed the limits of the models and have nonphysical derived quantities
(i.e. negative masses).}
\tablecomments{References:
1) Kraus et al. (in prep),
2) Luhman (2006),
3) Kenyon \& Hartmann (1995),
4) Leinert et al. (1993),
5) Tanner et al. (2007),
6) Luhman et al. (2000),
7) Luhman (2004),
8) Ghez et al. (1993),
9) White \& Basri (2003),
10) White et al. (in prep),
11) Herczeg \& Hillenbrand (2008),
12) Brice\~no et al. (1998),
13) Brice\~no et al. (2002),
14) Guieu et al. (2006),
15) Kraus et al. (2006),
16) Konopacky et al. (2007),
17) Luhman et al. (2003),
18) Strom \& Strom (1994),
19) White \& Hillenbrand (2004),
20) Luhman et al. (2006),
21) Hartigan et al. (1994),
22) Sartoretti et al. (1998),
23) Hartigan et al. (1995),
24) Simon et al. (1995),
25) Mart\'in et al. (2001),
26) Slesnick et al. (2006),
27) Brice\~no et al. (1999).}
\end{deluxetable*}

\clearpage
\end{landscape}
\end{LongTables}

\end{document}